\documentclass[aps,prb,twocolumn,floatfix,superscriptaddress]{revtex4-1}
\usepackage{bm}
\usepackage{epsf}
\usepackage{amssymb}
\usepackage{amsmath}
\usepackage{graphicx}
\usepackage{rotating}
\usepackage{epsfig}
\usepackage{psfrag}
\usepackage{amsmath}
\usepackage{hyperref}
\usepackage{subfigure}
\usepackage{color}
\usepackage{enumerate}

\usepackage{appendix}
\usepackage{chngcntr}

\renewcommand{\a}{\alpha}
\newcommand {\aplt} {\ {\raise-.5ex\hbox{$\buildrel<\over\sim$}}\ }

\newcommand{\glF}{\mathcal{F}}
\newcommand{\glFiso}{\mathcal{F}_{J}}
\newcommand{\glFaniso}{\mathcal{F}_{A}}
\newcommand{\glFDM}{\mathcal{F}_\mathrm{DM}}

\newcommand{\be}{\begin{equation}}
\newcommand{\ee}{\end{equation}}

\def\mb{\mathbf}
\def\mr{\mathrm}
\def\mc{\mathcal}

\begin{document}
\title{Skyrmions in Chiral Magnets with Rashba and Dresselhaus Spin-Orbit Coupling}

\author{James Rowland}
\affiliation{Department of Physics, The Ohio State University, Columbus, Ohio, 43210}
\author{Sumilan Banerjee}
\affiliation{Department of Condensed Matter Physics, Weizmann Institute of Science, Israel, 7610001}
\author{Mohit Randeria}
\affiliation{Department of Physics, The Ohio State University, Columbus, Ohio, 43210}

\date{\today}

\begin{abstract}
Skyrmions are topological spin textures of interest for fundamental science and applications.
Previous theoretical studies have focused on systems with broken bulk inversion symmetry, where skyrmions are stabilized by easy-axis anisotropy. 
We investigate here systems that break surface-inversion symmetry, in addition to possible broken bulk inversion. This leads to two distinct 
Dzyaloshinskii-Moriya (DM) terms with strengths $D_\perp$, arising from Rashba spin-orbit coupling (SOC), and $D_\parallel$ from Dresselhaus SOC. 
We show that skyrmions become progressively more stable with increasing $D_\perp/D_\parallel$, extending into the regime of easy-plane anisotropy. 
We find that the spin texture and topological charge density of skyrmions develops nontrivial spatial structure, with quantized topological charge in a unit cell given by a Chern number. Our results give a design principle for tuning Rashba SOC and magnetic anisotropy to stabilize skyrmions in thin films, surfaces, interfaces and bulk magnetic materials that break mirror symmetry.
\end{abstract}

\maketitle

Recently there has been a surge of interest in skyrmions in chiral magnetic 
materials~\cite{JMMMBogdanov1994,NatureRossler2006,NatureNanoNagaosa2013},
ranging from fundamental science to potential device applications.  A skyrmion is a spin texture 
characterized by a topological invariant that, in metallic magnets, gives rise to the topological Hall 
effect~\cite{PRLNeubauer2009,PRLLee2009} 
and may also have implications for non-Fermi liquid behavior~\cite{NatureRitz2013}.
The ability to write and erase individual skyrmions~\cite{ScienceRomming2013},
along with their topological stability, small size, and low depinning current density~\cite{ScienceJonietz2010},
paves the way for potential information storage and processing applications.

Experiments have focussed primarily on skyrmions in non-centrosymmetric crystals with broken bulk inversion symmetry,
e.g., metals like MnSi, FeGe and insulators like Cu$_2$OSeO$_3$. In these bulk materials, the skyrmion crystal (SkX)
phase is stable only in a very limited region of the magnetic field ($H$), temperature ($T$) phase 
diagram~\cite{ScienceMuhlbauer2009,NatureMatYu2011,ScienceSeki2012,Binz2006,PRBWilson2014}.
On the other hand, the skyrmion phase is found to be stable over a much wider region of $(T,H)$ in thin films of the same 
materials~\cite{PRLHuang2012,NanoLettTonomura2012,NatureMatYu2011,NatYu2010},
even extending down to $T\!=\!0$ in some cases~\cite{NatYu2010, PRLHuang2012}. 
(A class of two dimensional (2D) systems shows
atomic-scale skyrmions~\cite{NatPhysHeinze2011} arising from competing 
local interactions, distinct from the spin-orbit induced chiral interactions that we focus on here.)

A key question that we address is this paper is: How can we enhance the domain of stability of skyrmion spin textures?
We are motivated in part by the thin film experiments, and also by the possibility of chiral magnetism in new 2D systems like oxide interfaces~\cite{NaturePhysBanerjee2013,PRXBanerjee2014,PRLBalents2014}.
We show how the SkX become progressively more stable over ever larger regions in parameter space
of field $H$ and magnetic anisotropy $A$, as the effects of broken surface inversion dominate over those of
broken bulk inversion. The key parameter responsible for this behavior
is the ratio $D_\perp/D_\parallel$ of the strength of the chiral magnetic interaction arising from broken bulk inversion ($D_\parallel$) to that arising
from broken surface inversion ($D_\perp$); see Fig.~1.

We begin by summarizing our main results, which requires us to introduce some terminology.
We focus on magnets in which spin textures arise from the interplay between ferromagnetic exchange $J$
and the chiral Dzyaloshinskii-Moriya (DM) interaction $\mb{D}_{ij}\!\cdot\!(\mb{S}_i\!\times\!\mb{S}_j)$.  
Spin-orbit coupling (SOC) determines the magnitude of the $\mb{D}$ vector, while symmetry dictates its direction.  
Broken bulk inversion symmetry (${\bf r}\!\rightarrow\!-{\bf r}$) leads to the Dresselhaus DM term with
$\mb{D}_{ij} =D_\parallel\; \widehat{\mb{r}}_{ij}$, where $\widehat{\mb{r}}_{ij} = {\mb{r}}_{ij}/\vert{\mb{r}}_{ij}\vert$
with ${\mb{r}}_{ij}\!=\!\left({\mb{r}}_{i}\!-\!{\mb{r}}_{j}\right)$.
On the other hand, broken surface inversion or mirror symmetry 
($z\!\rightarrow\!-z$) leads to the Rashba DM term with $\mb{D}_{ij} =D_\perp(\widehat{\mb{z}}\times\widehat{\mb{r}}_{ij})$. 
In the limit of weak SOC, $D/J \ll 1$, where $D = (D_\parallel^2 + D_\perp^2)^{1/2}$, the length scale of 
spin textures is $(J/D) a \gg a$ (the microscopic lattice spacing) and we can work 
with a continuum ``Ginzburg-Landau'' field theory.

\begin{figure*}[!htb]
	\includegraphics[width=0.9\textwidth]{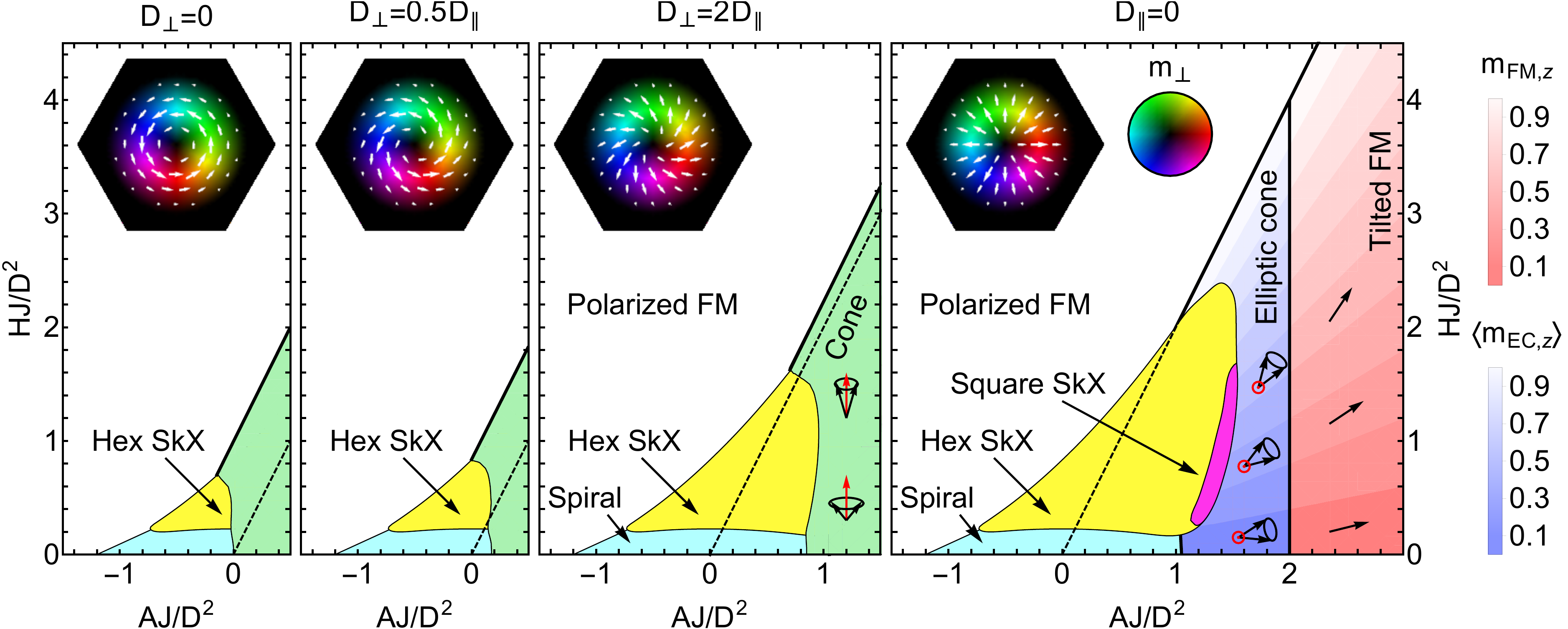}
	\caption{Phase diagrams as a function of $AJ/D^2$ and $HJ/D^2$ for four values of $D_\perp/D_\parallel$. 
	Easy-axis anisotropy corresponds to $A<0$ while easy-plane to $A>0$.
	The cone, elliptic cone, and tilted FM phases are 
	shown schematically, with the Q-vector shown in red and  the texture traced out by spins shown in black.
	The color bar on the right indicates $m_z$ for the elliptic cone and tilted FM phases in the $D_\parallel=0$ panel. 
	Insets: Unit cell in the hexagonal (Hex) skyrmion crystal (SkX) phase with white arrows indicating the projection of 
	magnetization on the $x$-$y$ plane. The colors indicates the magnitude and direction of the spin projection following 
	the convention of ref.~\onlinecite{NatureNanoNagaosa2013} indicated in the color wheel.  
      Thick lines denote continuous transitions, while thin lines indicate first-order phase transitions.
	\label{Fig1}
	}
\end{figure*}

We show in Fig.~1 the evolution of the $T\!=\!0$ phase diagram going from the pure Dresselhaus limit to the pure Rashba limit. 
Each phase diagram is plotted as a function of the (dimensionless) field $HJ/D^2$ and anisotropy $AJ/D^2$. Here $A>0$ ($A<0$) corresponds to 
easy-plane (easy-axis) anisotropy. Our main results are:

(1) As the Rashba $D_\perp$ is increased relative to the Dresselhaus $D_\parallel$, the spiral and skyrmion phases 
become increasingly more stable relative to the vertical cone phase, and penetrate into the easy-plane anisotropy side of
the phase diagram.

(2) With increasing $D_\perp/D_\parallel$, the textures change continuously from a Bloch-like spiral to
a Neel-like spiral. Correspondingly, the skyrmion helicity evolves with a vortex-like structure in the Dresselhaus limit to a hedgehog in the Rashba limit,
which is shown to impact the ferrotoroidic moment.

(3) In the pure Rashba limit, we find the largest domain of stability for the hexagonal skyrmion crystal. In addition we also find a small sliver of
stability for a square skyrmion lattice, together with an elliptic cone phase, distinct from the well-known vertical cone phase in the Dresselhaus limit.

(4) We see in Fig.~2 that in the Rashba limit the spin texture of the skyrmion and their topological charge density 
$\chi({\bf r})$ begins to show non-trivial spatial variations as one
changes anisotropy, but the total topological charge $N_{\rm sk}=\int d^2r\, \chi({\bf r})$ in each unit cell remains quantized, even when $\chi({\bf r})$
seems to ``fractionalize'' with positive and negative contributions within a unit cell.
 
(5) For $H > 2A$, one can have isolated skyrmions in a ferromagnetic (FM) background,
and their topological charge $N_{\rm sk}$ is quantized, as usual, by the homotopy group $\pi_2(S^2) = \mathbb{Z}$. 
For $H < 2A$, we find that skyrmions cannot exist as isolated objects, and 
$N_{\rm sk}$ must now be defined by the $\mathbb{Z}$ Chern number 
classifying maps from the SkX unit cell, a two-torus $T^2$ to $S^2$, the unit sphere in spin-space,
a definition that works for all values of $H/2A$.

\begin{figure*}[!htb]
	\includegraphics[width=0.8\textwidth]{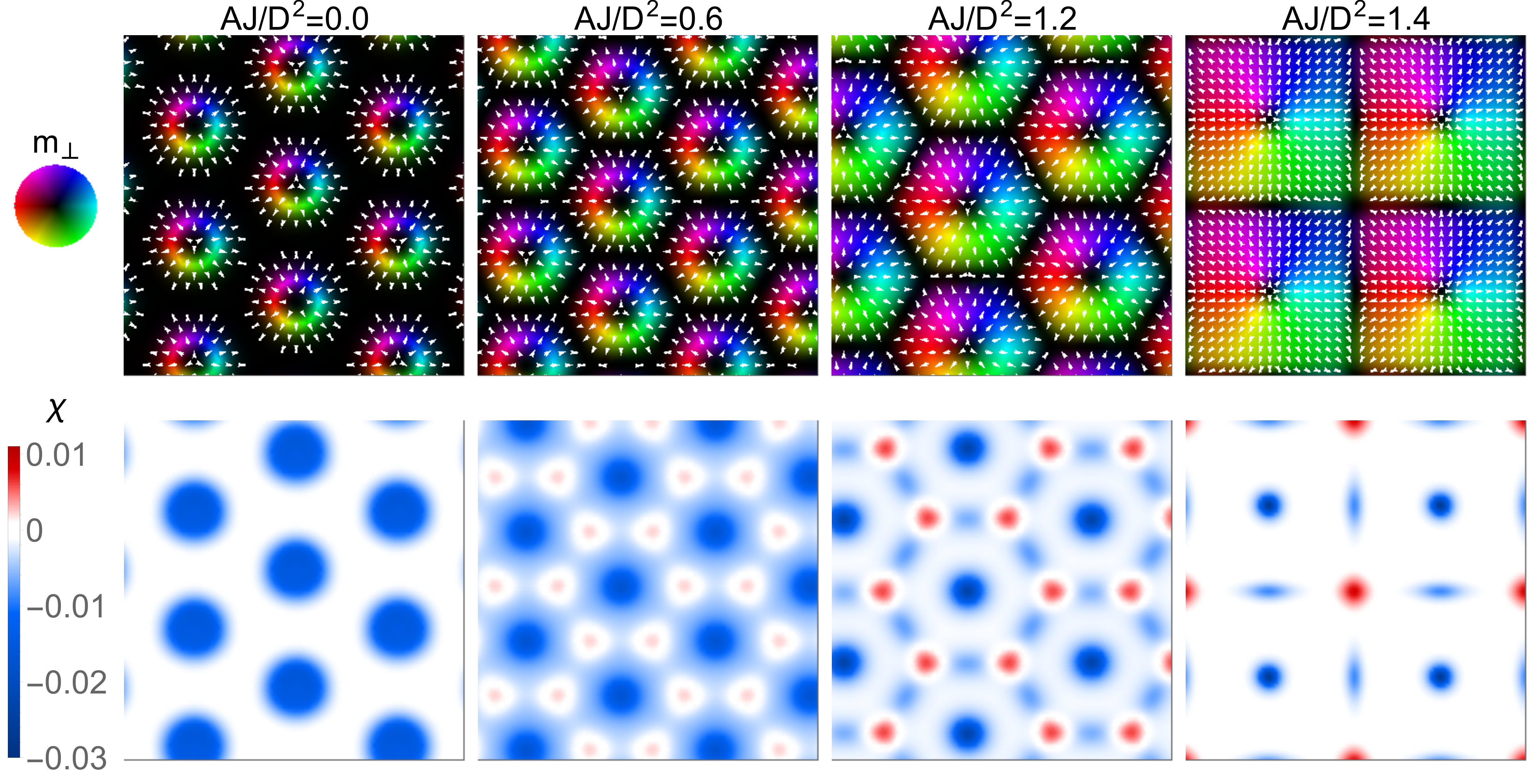}
	\caption{   
	Evolution of the spin texture $\mb{m}$ (top row) and the topological charge density $\chi$ (bottom row) 
	for four values of $AJ/D^2$ at fixed $HJ/D^2=0.7$ in the
	Rashba limit ($D_\parallel=0$).  White arrows indicate the projection of $\mb{m}$ into the $x$-$y$ plane. 
	The colors also indicates the magnitude and direction of the spin      
	projection following the convention of ref.~\onlinecite{NatureNanoNagaosa2013} indicated in the color wheel.  The development of nontrivial spatial variation in 
	$\chi({\bf r})$ is discussed in the text. Note, however, that in each case integral over a single unit cell $\int d^2{\bf r}\, \chi({\bf r}) = -1$.
	\label{Fig2}
	}
\end{figure*}

{\bf Free energy:}
We consider a continuum (free) energy functional $F[\mb{m}]=\int d^3r \mc{F}(\mb{m})$ 
with
\be\label{glF}
\glF=\glFiso+\glFDM+\glFaniso-H m_z
\ee
whose form is dictated by symmetry.
The isotropic exchange term $\mc{F}_J=(J/2)\sum_\a(\nabla m_\a)^2$ ($\a=x,y,z$) controls the gradient energy through stiffness $J$.
The DM contribution in the continuum 
\be\label{glFDM}
\glFDM = D\cos\beta\;\mb{m}\cdot (\nabla\times \mb{m}) +  D\sin\beta\;\mb{m}\cdot[(\hat{\mb{z}}\times\nabla)\times \mb{m}]
\ee
is the sum of the two terms discussed above. The
$D_\parallel =  D\cos\beta$ term arises from Dresselhaus SOC in the absence of bulk inversion and 
$D_\perp =  D\sin\beta$ from Rashba SOC with broken surface inversion.
The anisotropy term $\mc{F}_A=A m_z^2$ can be either easy-axis ($A<0$) or easy-plane ($A>0$). 
Several different mechanisms contribute to $A$, including single-ion and dipolar shape anisotropies. 
In addition, Rashba SOC naturally leads to an easy-plane, compass anisotropy $A_\perp \simeq D_\perp^2/2J$,
which is energetically comparable to the DM term~\cite{NaturePhysBanerjee2013,PRXBanerjee2014}.
We treat $A$ as a free, phenomenological parameter. 

We focus on $T\!=\!0$ where the local magnetization is constrained to have a fixed length $\mb{m}^2(\mb{r})=1$,
and it should be hardest to stabilize skyrmions; once $\vert\mb{m}(\mb{r})\vert$ can become smaller due to thermal fluctuations, 
skyrmions should be easier to stabilize.
It is convenient to scale all distances by the natural length scale $J/D$ (setting the microscopic $a=1$)
and scale the energy $\glF$ by $D^2/J$. All our results will be presented in terms the
three dimensionless parameters that describe $\glF$, namely field $HJ/D^2$, 
anisotropy $AJ/D^2$, and $\tan\beta = D_\perp/D_\parallel$.  

{\bf Phase Diagram:}
In Fig.~1, we show the evolution of the $(A,H)$ phase diagram as a function of $\tan\beta = D_\perp/D_\parallel$,
increasing from left to right. These results were obtained by minimizing the energy functional $\glFDM$ subject to $\mb{m}^2(\mb{r})=1$.
The energies of the fully polarized ferromagnet (FM), the tilted FM, and the vertical cone states can be determined analytically,
while the energies for the spiral, the skyrmion crystals and the elliptic cone state were
found by a numerical, conjugate gradient minimization approach. In all cases, the results were checked by semi-analytical 
variational calculations. 
Details of the methodology are described in the Supplementary Materials; here we focus on the results.

We begin with well known~\cite{PRBWilson2014} results in the Dresselhaus limit (left panel of Fig.~1), where
the hexagonal SkX and spiral phases are stable only in a small region with easy-axis anisotropy ($A\!\leq\!0$). 
The $A\!>\!0$ region is dominated by the vertical cone phase, where 
$\mb{m}_\mr{cone}(z)=(\cos\varphi(z)\sin\theta_0,\sin\varphi(z)\sin\theta_0,\cos\theta_0)$
with $\varphi(z)=D_\parallel z/J$ and $\cos\theta_0=H/[2A+D_\parallel^2/J]$. The phase boundary between the vertical cone and 
polarized FM is given by $H=2A + D_\parallel^2/J$.

We note a change of variables that greatly simplifies the 
analysis of skyrmion crystal and spiral phases. This transformation is useful when
$\mb{m}\!=\!\mb{m}(x,y)$ has no $z$-dependence (along the field). 
Using a rotation $R_z(-\beta)$ by an angle $-\beta$ about the $z$-axis, we define $\mb{n}(x,y) = R_z(-\beta)\mb{m}(x,y)$.
It is then easy to show that the two terms in (\ref{glFDM}) combine to give a pure Dresselhaus form 
$\glFDM = D \mb{n} \cdot (\nabla\times\mb{n})$.
The other terms in (\ref{glF}) are invariant under this transformation, and
thus $\glF$ greatly simplifies using the transformation $\mb{m}\!\rightarrow\!\mb{n}$.

We choose, without loss of generality, $\hat{{\bf x}}$ as the propagation direction for the spiral
of period $L$, so that $\mb{n}_\mr{sp}(x)=(0,\sin\theta(x),\cos\theta(x))$  with
$\mb{n}_\mr{sp}(x+L)=\mb{n}_\mr{sp}(x)$. (Note that this is {\it not} in general a single-$\mb{q}$ spiral). 
We minimize $\mathcal{F}$ to find the optimal $L$ 
and optimal {\it function} $\theta(x)$, which is a 1D minimization problem.
For the SkX, we first pick a unit cell, hexagonal or square. We then find its optimal
size and optimal texture $\mb{n}=\left(\cos\varphi\sin\theta,\sin\varphi\sin\theta,\cos\theta\right)$,
by solving a 2D minimization problem to determine $\varphi(x,y)$ and $\theta(x,y)$ within a unit cell, 
subject to periodic boundary conditions.
We calculate the optimal $\mb{n}$ and transform it back to the magnetization $\mb{m}$ at the end.

With increasing $D_\perp/D_\parallel$, we see that the SkX and spiral phases become more stable relative to the
vertical cone, and their region of stability extends into the easy-plane regime. To understand this,
consider increasing the Rashba $D_\perp$ keeping $D_\parallel$ fixed. The energy of cone $\mb{m}(z)$ depends only
on $D_\parallel$, and is unchanged as $D_\perp$ increases. In contrast, the SkX and spiral, with $\mb{m}= \mb{m}(x,y)$,
utilize the full $D=(D_\parallel^2+D_\perp^2)^{1/2}$ to lower their energy.

{\bf  Helicity and Ferrotoroidic Moment:}
We find that the spin textures smoothly evolve as a function of $D_\perp/D_\parallel$. The spiral continuously changes from 
a Bloch-like (helical) spiral in the Dresselhaus limit to a Neel-like (cycloid) spiral in the Rashba limit.
In between, the spins tumble around an axis 
at an angle $\beta = \tan^{-1}(D_\perp/D_\parallel)$ to the $\mb{q}$-vector of the spiral. Similarly
the skyrmions smoothly evolve from vortex-like textures in the Dresselhaus limit to hedgehogs in the Rashba limit, as seen
in the insets of Fig.~1. In fact, $\gamma=\pi/2-\beta$ is the ``helicity''~\cite{NatureNanoNagaosa2013} of the skyrmions.
 
Our results imply that $D_\perp/D_\parallel$ controls the helicity $\gamma$, where Rashba $D_\perp$ could be tunable by electric field
at an interface or by strain in a thin film. The ability to tune $\gamma$ could be important in several ways. There is 
a recent proposal to use helicity to manipulate the Josephson effect in a superconductor/magnetic-skyrmion/superconductor 
junctions~\cite{Yokoyama2015}. Another interesting phenomenon that depends on the helicity of 
skyrmions~\cite{PrivateCommunicationBatista} is the ``ferrotoroidic moment''
$\mb{t}=(1/2)\int d^3r[\mb{r}\times\mb{m}(\mb{r})]$ \cite{JPhysCondMatSpaldin2015,Castan2012}. 
We will show elsewhere that $\mb{t} = t_0\sin\gamma\, \widehat{\mb{z}}$ for the SkX.

{\bf Rashba limit:} Next we turn to the $D_\parallel=0$ results in the right panel of Fig.~1, where one has the maximum
regime of the stability for the spiral and the hexagonal SkX, in addition to a small region with a square lattice SkX
(first predicted in ref. \onlinecite{PRBBatista2015}), an elliptic cone phase and a tilted FM. This phase diagram improves upon all 
previous works~\cite{PRXBanerjee2014,PRLBalents2014,PRBBatista2015} as explained in detail in the Supplementary Material. 

The tilted FM, which spontaneously breaks the $U(1)$ symmetry of $\glF$ (in a field), has $m_z = H/2A$ and
exists in the regime $2A > H$ and $AJ/D_\perp^2 > 2$ for $D_\parallel=0$. 
We also see a new phase where the  
spins trace out a cone with an elliptic cross-section. The
elliptic cone axis makes an angle $\theta_0 = \cos^{-1}(H/2A)$ with $\hat{\bf z}$,
and the spatial variation of $\mb{m}$ is along a $\mb{q}$-vector in the $x$-$y$ plane.

The nature of various phase transitions is discussed in the Supplementary Materials. 
In Fig.~\ref{Fig1}, thick lines denote continuous while thin lines denote first-order transitions.
$(A=2,H=4)J/D^2$ is a Lifshitz point~\cite{Chaikin-Lubensky} 
at which a state without broken symmetry
(polarized FM) meets a broken symmetry (tilted FM) and 
a modulated (elliptic cone) phase.

Let us next consider deviations from pure Rashba limit to see how the extreme
right panel of Fig.~1 evolves into the $D_\parallel\neq 0$ phase diagrams.
As soon as one breaks bulk inversion, an infinitesimal $D_\parallel$ leads to
the tilted FM being overwhelmed by the vertical cone, which gains Dresselhaus DM energy.
On the other hand, the elliptic and vertical cone states compete for $D_\parallel\neq 0$
and for some small, but finite, $D_\parallel$ the vertical cone wins.

{\bf Spin textures and topological charge:}
There are interesting differences between the skyrmions for $H<2A$ and $H>2A$. 
($H=2A$ is marked as a dashed line in the phase diagrams of Fig.~1). First, let us look at
the spin textures. For SkX with $H>2A$, which have been the focus of all the past work, the spins at the boundary
of the unit cell (u.c.) are all up, parallel to the field. Hence one can think of isolated skyrmions in a fully polarized FM
background; see Fig.~2 left-most panels. It is the identification of the point at infinity in real space for an isolated skyrmion that lets us
define a map from $S^2 \rightarrow S^2$ and use the homotopy group
 $\pi_2(S^2) = \mathbb{Z}$ to characterize the topological charge or skyrmion number $N_\mr{sk}$.

In contrast, when $H<2A$, we find that the spins at the boundary are {\it not} all pointing up and the
only constraint is periodic boundary conditions on the u.c.; see Fig.~2.
There is no way to isolate this spin texture in a FM background.
We must now consider the map $\mb{r} \rightarrow \mb{m}(\mb{r})$ 
from the u.c., which is  a 2-torus $T^2$ to $S^2$ in spin space.
(Such maps are well known when $T^2$ represents a Brillouin zone in $\mb{k}$-space, but the mathematics is identical.)
This map is characterized by an integer Chern number $N_\mr{sk} = \int_{\rm u.c.} d^2\mb{r}\, \chi(\mb{r})$, where 
$\chi(\mb{r}) =  \mb{m} \cdot (\partial_x\mb{m}\times \partial_y \mb{m})/4\pi$ is the topological charge density.
In fact, one can use this definition of $N_\mr{sk}$ for all values of $H/2A$.

From the $A=0$ panel on the left side of Fig.~2, we see that $\chi(\mb{r})$ is concentrated near the center of the u.c. and it is always of the same sign, 
as it is for all $H>2A$. With increasing $A$, once $H < 2A$, we see that $\chi(\mb{r})$ begins to spread out and even changes sign within the u.c.
In the square SkX phase $\chi$ is again concentrated, but this time in regions near the center and the edges of each u.c. 
along with regions of opposite sign at the u.c.~corners. For $H\aplt 2A$, the spin textures in the SkX phases are 
essentially composed of vortices and anti-vortices. Nevertheless, the Chern number argument shows that the total topological charge 
in each u.c. is an integer; $N_\mr{sk} = -1$ in all the panels of Fig.~2.

{\bf Discussion:} 
Previous theories on understanding the increased stability on skyrmions in thin films of non-centrosymmetric 
materials~\cite{PRLHuang2012,NanoLettTonomura2012,NatureMatYu2011,NatYu2010} focussed primarily on 
the changes in {\it uniaxial} magnetocrystalline anisotropy~\cite{Wilson2012,PRBWilson2014,Karhu2012} with thickness,
or on finite-size effects~\cite{Rybakov2013,Rybakov2015}. 
In fact, the latter can give rise to spin-textures more complicated than skyrmion crystals, with variations in all three directions. 
However, none of these theories take into account the role of broken surface inversion and Rashba SOC. As we have shown here, a non-zero Rashba
$D_\bot$ leads to a greatly enhanced stability of the SkX phase, particularly for {\it easy-plane} anisotropy, while
at the same time giving a handle on the helicity of skyrmions with interesting internal structure.

We note that the phase diagrams in Fig.~\ref{Fig1} apply to all systems with broken mirror symmetry, with or without bulk inversion. Mirror symmetry can be broken by certain crystal structures in bulk materials, 
by strain~\cite{Kato2004} in thin films, or by electric fields at interfaces. For systems with $D_\parallel=0$ the 
vertical cone phase, which dominates much of the phase diagram for $D_\parallel\neq0$, simply does not exist. 
After our paper was written, we became aware of the very recent observation~\cite{Kezsmarki2015} of hedgehog-like 
skyrmions in the magnetic semiconductor GaV$_4$S$_8$, a polar material with broken mirror symmetry that is
dominated by Rashba SOC. Skyrmions are, however, stabilized in this material only at finite temperature due to the 
large easy-axis anisotropy.

In conclusion, we have made a comprehensive study of the $T\!=\!0$ phases, with a focus on skyrmion crystals in chiral magnets with two 
distinct DM terms. $D_\perp$, arises from Rashba SOC and broken surface inversion, while 
$D_\parallel$ comes from Dresselhaus SOC and broken bulk inversion symmetry.
We predict that increasing the Rashba SOC, via strain or electric field, and tuning magnetic anisotropy towards the easy-plane side will
greatly help stabilize skyrmion phases in thin films, surfaces, and interface magnetism. 
Our results are very general, based on a continuum ``Ginzburg-Landau'' energy functional whose form is dictated by symmetry.
We hope that it will motivate ab-initio density functional theory calculations of the relevant phenomenological parameters entering 
our theory and an experimental investigations of skyrmions in Rashba systems.

{\bf Acknowledgments}:
We thank C. Batista, S. Lin and N. Nagaosa for useful discussions. MR acknowledges support from NSF DMR-1410364. 
JR was supported by an NSF Graduate Research Fellowship, and by the CEM, 
an NSF MRSEC, under grant DMR-1420451.

\vfill\eject
\bigskip
\centerline{\bf APPENDICES}

\tableofcontents

\begin{appendix}
\counterwithin{figure}{section}


\section{Continuum Free Energy}\label{sec.Continuum Model}
In this appendix we first introduce the continuum free energy functional that we use to model the spin textures
in a chiral magnet.
The free energy for a magnetic system with broken bulk inversion and mirror symmetries is $F[\mb{m}(\mb{r})]=\int d^3r\mc{F(\mb{m(\mb{r})})}$ where
\begin{eqnarray}\label{eqn.free energy unscaled}
\mc{F}(\mb{m}(\mb{r})) & = & (J/2)(\nabla \mb{m})^2 \\
 &  & + {D}_\parallel\, \mb{m}\cdot\left(\nabla\times\mb{m}\right) \nonumber \\
 &  & + {D}_\bot\, \mb{m}\cdot\left(\left(\hat{\mb{z}}\times\nabla\right)\times\mb{m}\right) \nonumber \\
 &  & + A m_z^2 - H m_z. \nonumber
\end{eqnarray}
and $(\nabla\mb{m})^2$ is shorthand for $\sum_{i,\alpha}(\partial_i m_\alpha)^2$. The $z$-axis is the axis of broken mirror symmetry. 
Here $J$ is the exchange stiffness, $\mb{D}_{ij}= {D}_\parallel\hat{\mb{r}}_{ij}+{D}_\bot\hat{\mb{z}}\times\hat{\mb{r}}_{ij}$ is the DM vector, $A$ is
the magnetic anisotropy, and $H$ is the field.
We measure all the lengths in units of microscopic lattice spacing $a$, which we set to unity, so that
$J$, $D_\parallel$, $D_\bot$, $A$ and $H$ all have units of energy.
We are interested in low temperature behavior so we ignore fluctuations in the magnitude of the 
local magnetization $\mb{m}$ and impose the constraint that $|\mb{m}(\mb{r})|=1$.

In our free energy functional we take the normal to the 
plane in which mirror symmetry is broken and the easy-axis direction to be the same,
namely $\hat{\bf z}$. For simplicity, we also choose the external field to be along the same direction.
One can, of course, imagine more general situations in which these directions are not all the same,
but the ``simple'' case treated here has sufficient complexity that it must be thoroughly 
investigated first.
 
Next we define parameters $D$ and $\beta$ such that 
\begin{equation}\label{eqn.def.D.beta}
D_\parallel=D\cos\beta \ \ \ {\rm and} \ \ \  D_\bot=D\sin\beta. 
\end{equation}
It is convenient to rewrite (\ref{eqn.free energy unscaled}) using the natural
energy and length scales in the problem. 
We measure energies in units of $D^2/J$ and lengths in units of $J/D$. In scaled variables 
the free energy density is given by
\begin{eqnarray}\label{eqn.free energy scaled}
\mc{F}(\mb{m}(\mb{r})) & = & (1/2)(\nabla \mb{m})^2 \\
 &  & + \mb{m}\cdot\left([\cos\beta\, \nabla+\sin\beta\, \hat{\mb{z}}\times\nabla]\times\mb{m}\right) \nonumber \\
 &  & + A m_z^2 - H m_z, \nonumber
\end{eqnarray}
which depends on three dimensionless
variables $AJ/D^2$, $HJ/D^2$ and $\beta=\tan^{-1}({D}_\bot/{D}_\parallel)$. 
In the main paper, we explicitly write the anisotropy and field as $AJ/D^2$ and $HJ/D^2$,
but in the Appendices we simplify notation and denote them as just $A$ and $H$.
(The total energy $F$
depends on an inconsequential overall factor of $(J/D)^3$ coming from the integration over the volume.)

Next we briefly discuss the phases that we find as a function of $A$, $H$ and $\beta$. 
The two ferromagnetic (FM) phases -- fully polarized and tilted -- are 
states with no spatial variations. The (vertical) cone phase is a non-coplanar
state which has only $z$-variations, along the magnetic field direction, 
so that $\mb{m} = \mb{m}(z)$. These states
can be treated analytically, as discussed in Appendix \ref{sec.Ferromagnetic and Cone Phases}.

The spiral and skyrmion phases have a local magnetization of the form
$\mb{m} = \mb{m}(x,y)$, as does the elliptic cone phase. 
Specifically, the spiral has spatial 
variation along a single direction, say $x$, in the plane perpendicular to the field.
The SkX phases have magnetic texture varying in both $x$ and $y$.
We use numerical methods for the analysis of phases with $\mb{m} = \mb{m}(x,y)$ as described
in Appendix~\ref{sec.Numerical Methods}.
Note that we do not consider states where ${\bf m}$ has non-trivial variations in all three coordinates;
see, e.g., ref.~\onlinecite{Rybakov2015}.

{\bf Rotation:} 
We next give the details of a transformation (introduced in the main text) that greatly simplifies
the analysis for states where ${\bf m} = {\bf m}(x,y)$. 
We make the rotation
\begin{equation}\label{eqn.rotation}
\mb{n}=\left(\begin{array}{rrr}
\cos\beta & \sin\beta & 0 \\
-\sin\beta & \cos\beta & 0 \\
0 & 0 & 1
\end{array}\right)\mb{m} \equiv R_z(-\beta)\, \mb{m}.
\end{equation}
where $\beta = \tan^{-1}(D_\bot/D_\parallel)$. Expressed in terms of $\mb{m}$
the free energy density (\ref{eqn.free energy scaled}) simplifies to
\begin{equation}\label{eqn.free energy rotated}
\mc{F}=(1/2)(\nabla\mb{n})^2+\mb{n}\cdot(\nabla\times\mb{n})+An_z^2-Hn_z,
\end{equation}
{\it provided} that $\mb{m}$,  and thus $\mb{n}$, depends only on $x$ and $y$, but not on $z$. 
This result (\ref{eqn.free energy rotated}) has the same form as the free energy density 
in the pure Dresselhaus limit. After solving the problem in the $\mb{n}$ representation,
we must transform back to $\mb{m}$ to find the actual spin texture.

It is easy to see that the the exchange term is invariant under $\mb{m} \to \mb{n}$, 
i.e., $(\nabla\mb{m})^2\to(\nabla\mb{n})^2$, as are the anisotropy term and Zeeman term coupling to $H$.
The only term that has a nontrivial transformation is DM term in (\ref{eqn.free energy scaled}).  
We symbolically write the DM term as as $\mb{m}\cdot\left({\cal  D}\times\mb{m}\right)$ where
 \begin{eqnarray} 
   {\cal D} &\equiv& \cos\beta\, \nabla+\sin\beta\, (\hat{\mb{z}}\times\nabla) \nonumber\\
    &=& \left(\begin{array}{ccc}
           \cos\beta\partial_x-\sin\beta\partial_y \\
           \sin\beta\partial_x+\cos\beta\partial_y \\
           0
        \end{array}\right).
\end{eqnarray}  
Here we set  ${\cal D}_z  = \cos\beta\partial_z \to 0$, because we focus on textures that have 
no $z$-variation, as already stated above. A straightforward calculation,
using $\mb{m} = R_z(\beta)\, \mb{n}$, then allows us to derive
\begin{equation}
\mb{m}\cdot\left({\cal  D}\times\mb{m}\right) = \mb{n}\cdot(\nabla\times\mb{n})
\end{equation}
which in turn leads to eq.~(\ref{eqn.free energy rotated}).

There is a slick way to obtain this same result by noting that the Free energy is invariant under 
a combined rotation in spin-space and in real space about the $z$-axis. 
(A combined spin and spatial rotation is needed because of SOC. 
and the $z$-axis is singled out by broken mirror symmetry). In fact, it is simple
to proceed with the general case where we retain ${\cal D}_z  = \cos\beta\partial_z$.
The transformation $R_z(-\beta)$ of eq.~(\ref{eqn.rotation}) acts {\it only} in spin-space.
Thus to write the free energy in terms of $\mb{n}$, we need to also rotate
$\nabla$ in real space, so that ${\cal D}$ transforms to
\begin{equation}
\left(\begin{array}{lll}
           \cos\beta[\cos\beta\partial_x-\sin\beta\partial_y]+\sin\beta[\sin\beta\partial_x+\cos\beta\partial_y]\\
          -\sin\beta[\cos\beta\partial_x-\sin\beta\partial_y]+\cos\beta[\sin\beta\partial_x+\cos\beta\partial_y]\\
           \cos\beta\partial_z
        \end{array}\right). \nonumber
\end{equation}
This can be simply written as 
\begin{equation}
{\cal D} \to \left[\nabla-(1-\cos\beta)\partial_z\hat{\mb{z}}\right].
\end{equation}
Thus, for any magnetic texture $\mb{m}(x,y,z)$, the DM term can we written in general as 
\begin{equation}
+ \mb{n}\cdot(\nabla\times\mb{n}) - (1-\cos\beta) \mb{n}\cdot(\partial_z(\hat{\mb{z}}\times\mb{n})) 
\end{equation}
For a magnetic texture $\mb{m}(x,y)$ that does not vary along the $z$-axis the $z$-derivative terms vanish 
and this result simplifies to (\ref{eqn.free energy rotated}) derived above.

\section{Ferromagnetic and Cone Phases}\label{sec.Ferromagnetic and Cone Phases}
In this appendix (\ref{sec.Ferromagnetic and Cone Phases}) we consider phases which can be treated analytically: the polarized ferromagnet (FM), tilted FM and the vertical cone phase.

{\bf Ferromagnets:} The free energy density for a FM state is
\begin{equation}\label{eqn.FM free energy}
\mc{F}=Am_z^2-Hm_z.
\end{equation}
Minimizing the free energy is easy in this case; we need to solve
\begin{equation}
0=\frac{\partial\mc{F}}{\partial m_z}=2Am_z-H
\end{equation}
along with the constraint $|\mb{m}|^2=1$. The solution is
\begin{equation}\label{eqn.FM solution}
m_z^*=\left\{\begin{array}{ll}
1 & H\ge 2A \\
\frac{H}{2A} & H<2A

\end{array}\right.\Rightarrow\mc{F}^*=\left\{\begin{array}{ll}
A-H & H\ge 2A \\
-\frac{H^2}{4A} & H<2A
\end{array} \right. .
\end{equation}
We call the solution with $m_z^*=1$ a polarized FM since the magnetization is aligned with the magnetic field. The solution with $m_z^*<1$ we call tilted FM since the magnetization is tilted away from the magnetic field. Unlike the polarized FM, the tilted FM spontaneously breaks the $U(1)$ symmetry of spin rotation around $z$ axis in \ref{eqn.FM free energy}.

\begin{figure}
\includegraphics[scale=0.25]{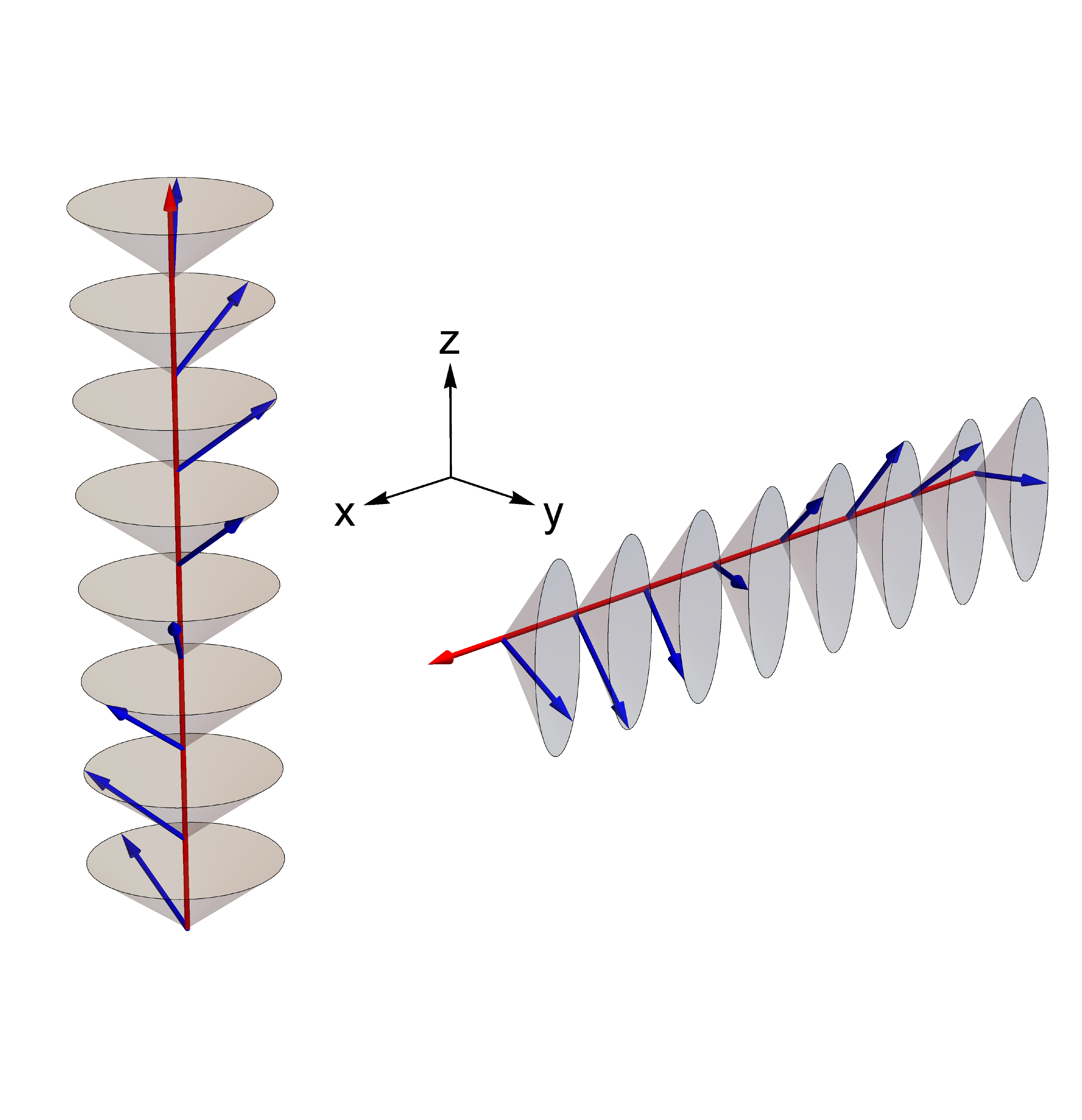}
\caption{\label{fig.cone illustration} Illustration of the vertical cone phase, with $\mb{q}$-vector (red arrow) along the $z$-axis, and the elliptic cone phase. In the elliptic cone phase the magnetization traces out an elliptic cone, i.e., the cross section is an ellipse rather than a circle. The elliptic cone phase shown here is for the $D_\parallel=0$ limit. If the cone height is decreased so that the spins lie in the $x$-$z$ plane the configuration becomes a cycloid (Neel-like spiral).}
\end{figure}

{\bf Cone:} Another simple class of magnetic states are textures $\mb{m}(z)$ that do not break translational symmetry in the $x$-$y$ plane. We will find that the optimum configuration is a cone texture. This texture is called a cone because the magnetic moments trace out a cone as a function of $z$; see the illustration in figure \ref{fig.cone illustration}.

For textures $\mb{m}(z)$ with translation symmetry in the $x$-$y$ plane the Rashba term in (\ref{eqn.free energy scaled}), with strength $\sin\beta$, does not contribute to the free energy. To implement the constraint $|\mb{m}(z)|=1$ we define angular variables $\theta(z)$ and $\phi(z)$ such that
\begin{equation}
\mb{m}(z)=(\cos\phi\sin\theta,\sin\phi\sin\theta,\cos\theta).
\end{equation}
In terms of $\theta(z)$ and $\phi(z)$ the free energy density is
\begin{eqnarray}\label{eqn.Cone free energy integral}
\mathcal{F}(\theta(z),\phi(z))&=&\frac{1}{2}(\theta')^2+\frac{1}{2}(\phi')^2\sin^2\theta-\phi'\cos\beta\sin^2\theta  \nonumber \\
 & & +A\cos^2\theta-H\cos\theta.
\end{eqnarray}
The Euler-Lagrange equation for $\phi(z)$,
\begin{equation}
0=\frac{\partial}{\partial z}\frac{\partial\mc{F}}{\partial \phi'}=\frac{\partial}{\partial z}(\phi'\sin^2\theta-\cos\beta\sin^2\theta),
\end{equation}
can be integrated with the result
\begin{equation}\label{eqn.phi equation integrated}
\phi'\sin^2\theta-\cos\beta\sin^2\theta=C.
\end{equation}
Thus the free energy density can be written
\begin{equation}
\mc{F}(\theta(z))=\frac{1}{2}(\theta')^2+f(\theta)
\end{equation}
where $f$ does not depend on $\theta'$. If $\theta^*$ is a minimum of $f$ then $f(\theta(z))\geq f(\theta^*)$ for all $z$. It is clear that $\theta(z)=\theta^*$ is an extremum for the free energy obtained from (\ref{eqn.Cone free energy integral}). Given that a constant $\theta(z)=\theta^*$ minimizes the free energy it is easy to find the optimum value $\theta^*$ and the extremal function $\phi$ which is 
\begin{equation}\label{cone.phi}
\phi^*(z)=z\cos\beta. 
\end{equation}
There are two solutions for $\theta^*$. The first solution is $\theta^*=0$ which is a ferromagnetic solution with energy $F=A-H$. The second solution, with 
\begin{equation}\label{cone.theta}
\theta^*=\cos^{-1}(H/(2A+\cos^2\beta)),
\end{equation}
is called a cone texture.  
To make contact with the main text, we must recall that lengths are
measured in units of $J/D$, so that eq.~(\ref{cone.phi}) becomes 
$\phi^*(z)\to z(D/J)(D_\parallel/D) = z D_\parallel/J$, and energies $A$ and $H$ in units of $D^2/J$, so that 
$\theta^* \to \cos^{-1}[H/(2A+D_\parallel^2/J)]$.

We will occasionally refer to the cone as a vertical cone to distinguish it from the elliptic cone found in Appendices \ref{sec.Numerical Methods} and \ref{sec.Variational Solution}. The elliptic cone varies in the $x$-$y$ plane while the vertical cone varies along the $z$-axis. An illustration of the vertical and elliptic cone textures is given in figure \ref{fig.cone illustration}. In the Rashba limit, $D_\parallel=0$, the cone phase is not stable for any values of $H$ and $A$. For finite $D_\parallel$ the tilted FM phase is not stable anywhere and the vertical cone takes its place in the phase diagram.

\section{Numerical Methods}\label{sec.Numerical Methods}
In appendix \ref{sec.Numerical Methods} we consider phases that cannot be treated analytically: the spiral, elliptic cone, square skyrmion crystal and hexagonal skyrmion crystal phases. For these phases the free energy density needs to be integrated and minimized numerically; we use conjugate gradient minimization to achieve this. All of these phases are of the form $\mb{m}(x,y)$ so we use the transformed free energy density (\ref{eqn.free energy rotated}) and we will refer to $\mb{n}(x,y)$ as the texture in the spin rotated frame. To implement numerical integration of the free energy some boundary conditions need to be specified. We use three boundary conditions: periodic along the $x$-axis, periodic with square symmetry and periodic with hexagonal symmetry. 

For each boundary condition the minimization procedure is very similar: convert the function $\mb{n}(x,y)$ to a vector $\mb{s}(i,j)$, write the integral as a sum, then minimize to find the optimum vector. We discuss this procedure in detail only for the simplest case (periodic along the $x$-axis) and we discuss the important differences for the other two cases.

{\bf Spiral and elliptic cone:} First we consider textures which are periodic along a single axis; here we choose the $x$ axis without loss of generality. The periodic boundary condition means that the texture $\mb{n}(x)$ satisfies $\mb{n}(x)=\mb{n}(x+L)$. Since the free energy density is uniform along the $y$ and $z$ axes the free energy is just a one-dimensional integral along the $x$ axis. 

To evaluate this integral numerically we can convert it to a sum. This involves discretizing the function $\mb{n}(x)$; let $\mb{s}(i)=\mb{n}((i-1)\Delta x)$ where $\Delta x=L/(N-1)$. Then $\mb{s}(i+N)=\mb{s}(i)$ is the periodic boundary condition. Next we replace all derivatives in the free energy density with finite differences, i.e., $\partial_x\mb{n}(x)\to(\mb{s}(i+1)-\mb{s}(i))/\Delta x$ for the point $x=(i-1)\Delta x$. Lastly we replace the integral with a sum ($\int_0^Ldx\to\Delta x\sum_{i=1}^N$). At this point the free energy can be computed given a configuration $\mb{s}(i)$ and this will be a good approximation to the true free energy when $N$ is large, i.e., $F(\mb{s})\approx F[\mb{n}]$ is a good approximation when $N$ is large.

We also have the constraint $|\mb{n}(x)|=1$. In terms of $\mb{s}$ the constraint is $|\mb{s}(i)|=1$. We impose the constraint by introducing angular variables, $\theta(i)$ and $\phi(i)$, at each site so that $\mb{s}(i)=(\cos\phi\sin\theta,\sin\phi\sin\theta,\cos\theta)$. Now we can think of $F$ as a function of the $2N$ component vector $\{\theta(1),...,\theta(N),\phi(1),...,\phi(N)\}=\{\bm{\theta},\bm{\phi}\}=\mb{p}$. We can minimize the vector function $F(\mb{p})$ using conjugate gradient minimization. Conjugate gradient minimization accepts as its input a function, in this case $F$, the gradient of that function $\nabla_\mb{p}F$, and an initial vector, call it $\mb{p}_0$; the output of conjugate gradient minimization is a local minimum $\mb{p}^*$. We choose single-$\mb{q}$ spirals $\mb{m}(x)=(\cos qx,\sin qx,0)$ for our initial condition where $q=2\pi/L$. This choice of initial condition does not limit the output of conjugate gradient minimization to single-$\mb{q}$ spirals; the output has a more complicated Fourier space structure in general.

Note that the function $F(\mb{p})$ must be a real valued function of $\mb{p}$. In particular we need to specify $A$, $H$ and the periodicity $L$ for $F$ to be a real valued function. Note that $L$ is also a variational parameter in this case. In order to find the optimum value of $L$ we include $L$ in the conjugate gradient minimization procedure, i.e., we think of $F(\mb{p},L)$ as a function of the $2N+1$ component vector $\{\mb{p},L\}$. We then sweep across the phase diagram and find the optimum configuration for each value of $-1.6<A<3.1$ and $0<H<4.1$ in steps of $0.1$ along both axes.

The results of this minimization procedure are incorporated into the phase diagram in Fig. 1 of the main text. The phases that we find are spiral phases and elliptic cone phases. Since we minimize the simplified free energy (\ref{eqn.free energy rotated}) the spirals we find are all Bloch-like spirals. These textures can be expressed in terms of a single function $\theta(x)$ where $\mb{n}(x)=(0,\sin\theta(x),\cos\theta(x))$. The textures obtained in the lab frame are related to these texture by a rotation by $\beta$ about the $z$-axis as discussed in Appendix \ref{sec.Continuum Model}, i.e., $\mb{m}(x)=(\sin\beta\sin\theta,\cos\beta\sin\theta,\cos\theta)$. 

The other phase we find is the elliptic cone. This phase is more complicated than the spiral phase. Where the spiral can be described in terms of a single function, $\theta(x)$, the elliptic cone requires two functions in general, i.e., $\theta(x)$ and $\phi(x)$. As a consequence of the $\phi(x)$ dependence all components of the magnetic moment vary as a function of $x$, not just the $y$ and $z$-components like the spiral. 
An illustration of the elliptic cone is given in Fig.~\ref{fig.cone illustration} and in Appendix \ref{sec.Variational Solution} we discuss a variational state that captures the physics of the elliptic cone.

\begin{figure}
\includegraphics[scale=0.4]{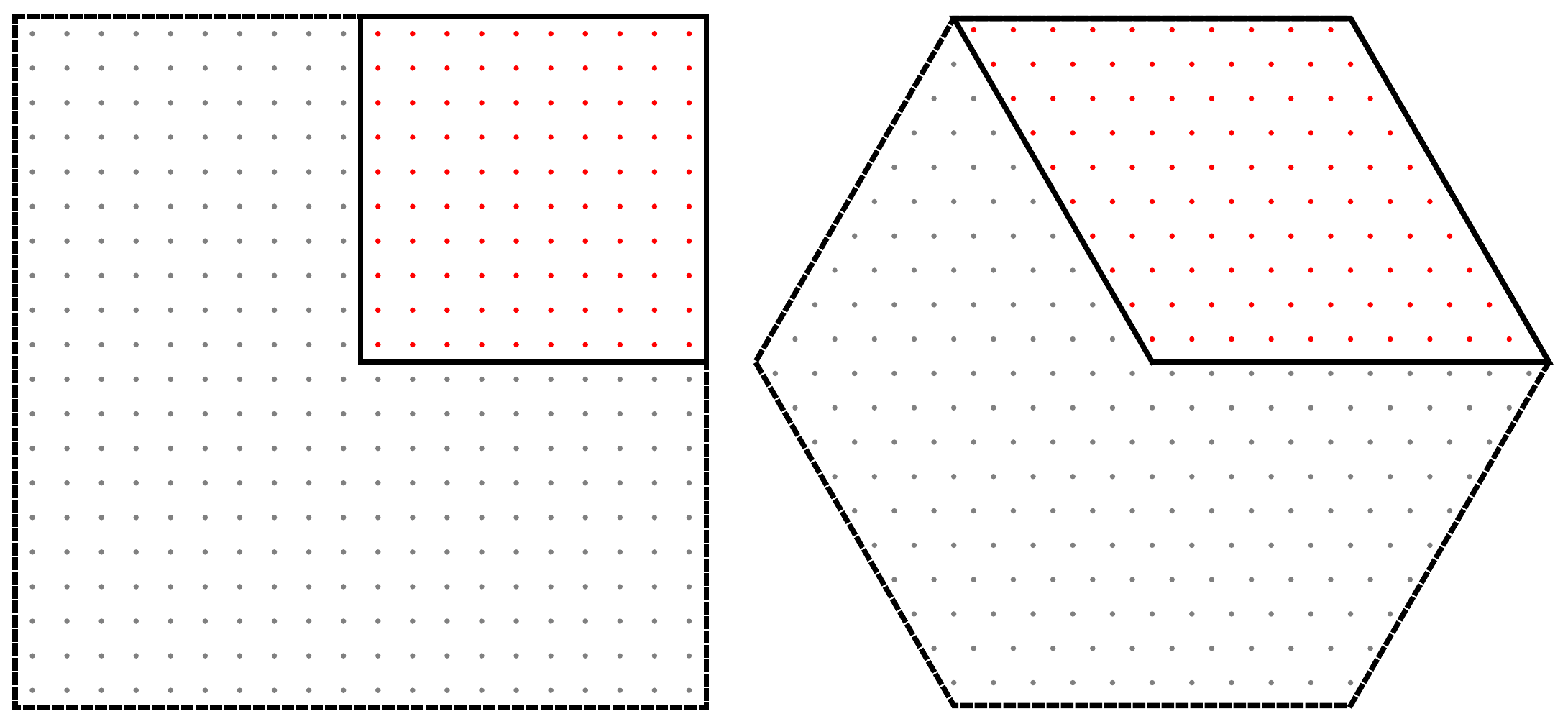}
\caption{\label{fig.unit cell illustration} Illustration of square (left) and hexagonal (right) unit cells. The dark lines indicate the region of independent spins (indicated by red dots). Spins in the dashed region (gray dots) are determined by $C_4$ symmetry for the square system and $C_3$ symmetry for the hexagonal system.}
\end{figure}

{\bf Square skyrmion crystal:} Next we consider textures that are periodic along both axes, i.e., $\mb{n}(x,y)=\mb{n}(x+L,y)=\mb{n}(x,y+L)$. We further restrict ourselves to textures with $C_4$ symmetry and we use this symmetry to reduce the number of points in the unit cell by a factor of 4 (see Fig. \ref{fig.unit cell illustration} for an illustration). We convert the free energy integral over the square unit cell to a sum and we convert derivatives to finite differences. Here we need a $2N\times N$ component vector $\mb{p}=\{\theta(1,1),...,\theta(1,N),...,\theta(N,N),\phi(1,1),...,\phi(N,N)\}$ to describe a given magnetic texture $\mb{n}(x,y)$. The unit cell has $8N\times N$ sites. We need to specify the function, $F$, its gradient, $\nabla_\mb{p}F$, and an initial vector $\mb{p}_0$. We use a single-$\mb{q}$ vortex-like skyrmion as our initial configuration, i.e.,
\begin{equation}
\mb{n}(x,y)=[-(y/r)\hat{\mb{x}}+(x/r)\hat{\mb{y}}]\sin (qr)+\hat{\mb{z}}\cos (qr),
\end{equation}
where $r = \sqrt{x^2 + y^2}$ and we use $q=\pi/L$. 
Note that this configuration is actually a square skyrmion crystal since we are applying square periodic boundary conditions. As we have already stressed, the simple form of the initial configuration does not limit the form of the output of conjugate gradient minimization.

To eliminate discretization error and approach the continuum limit, we extrapolate our results to the $N\to\infty$ limit to obtain the phase diagram in Fig. 1 of the main text. We use a polynomial fit of the energy vs $\Delta x$ for $N=40,...,60$ and we find that the $\Delta x=0$ 
limit 
is the same whether we fit to a polynomial of degree 3, 4 or 5. This gives us confidence that we are sufficiently close to the $N\to\infty$ limit that extrapolation is valid.

The optimum configurations we find using conjugate gradient minimization are square skyrmion crystals. The topological charge density $\mb{n}\cdot(\partial_x\mb{n}\times\partial_y\mb{n})$ is not concentrated at the center of each unit cell like it is for skyrmions found in the Dresselhaus limit. The topological charge density is concentrated at four points in the unit cell: the center, corners and edges of the unit cell (see Fig. 2 in the main text) with opposite sign at the corners. Such configurations have been referred to as meron crystals \cite{PRBBatista2015}; nevertheless, the skyrmion number for this phase is quantized in integer units by
the Chern number argument given in the main text. The square skyrmion crystal phase is stable in a small pocket of the $D_\parallel=0$ phase diagram.

{\bf Hexagonal skyrmion crystal} Lastly we consider textures with hexagonal symmetry, i.e., $\mb{x}(x,y)=\mb{m}(x+a_1,y+a_2)=\mb{m}(x+b_1,y+b_2)$ where $(a_1,a_2)$ and $(b_1,b_2)$ are lattice vectors for a regular triangular lattice with lattice spacing $L$ and we also assume $C_3$ symmetry. In figure \ref{fig.unit cell illustration} we illustrate how we use the $C_3$ symmetry to reduce the number of points in our simulation by a factor of 3. Besides the symmetry of the problem, the minimization follows exactly as for the square symmetry case.

The optimum configurations we find using conjugate gradient minimization are hexagonal skyrmion crystals. The hexagonal crystal has a much large range of stability than the square crystal. Above the line $H=2A$ the hexagonal crystal has topological charge density concentrated in the center of each unit cell and the topological charge density is essentially zero throughout the unit cell. All skyrmions found in the Dresselhaus limit are of this type. Below the $H=2A$ line the topological charge density becomes less dense at the center of the unit cell and the topological charge is smeared across the unit cell. Close to the square crystal phase the topological charge density gathers at the center, corners and edges of the unit cell, with the skyrmion number still remaining quantized.

\section{Variational Solution}\label{sec.Variational Solution}
In appendix \ref{sec.Variational Solution} we discuss a variational state that can be integrated exactly and captures the physics of the polarized FM, tilted FM, spiral, elliptic cone and vertical cone phases. The main point of this variational state is to gain analytical insight into the new elliptic cone phase; in particular, this variational state confirms the phase boundary between the elliptic cone and the tilted FM is $A=2$ and the phase boundary between the elliptic cone and the polarized FM is $H=2A$. 

Consider the spin-texture
\begin{equation}\label{eqn.cone variation}
\mb{m}(\mb{r})= \hat{\mb{e}}_1 m_1\cos\mb{q}\cdot\mb{r}+\hat{\mb{e}}_2 m_2\sin\mb{q}\cdot\mb{r}+\hat{\mb{e}}_3 m_3(\mb{r})
\end{equation}
where $\hat{\mb{e}}_1$, $\hat{\mb{e}}_2$ and $\hat{\mb{e}}_3$ are orthonormal vectors. The function $m_3(\mb{r})=\sqrt{1-(m_1\cos\mb{q}\cdot\mb{r})^2-(m_2\sin\mb{q}\cdot\mb{r})^2}$ is defined to enforce the normalization  $|\mb{m}(\mb{r})|=1$ at every point. We must restrict $m_1$ and $m_2$ to the range $-1<m_1,m_2<1$ to make sure $m_3$ is a real number. Notice that this state varies along the $\hat{\mb{q}}$ axis and is uniform perpendicular to this axis.

{\bf Limits:} Here we discuss various limits of the variational state. The simplest case is the limit $m_1=m_2=0$ where the variational state reduces to $\mb{m}=\hat{\mb{e}}_3$ which is just a ferromagnetic configuration, either polarized ($\hat{\mb{e}}_3=\hat{\mb{z}}$) or tilted.

Another simple limit is $m_1=m_2=1$. In this case the variational state becomes $\mb{m}=\hat{\mb{e}}_1\cos\mb{q}\cdot\mb{r}+\hat{\mb{e}}_2\sin\mb{q}\cdot\mb{r}$ which is a spiral state. Notice that this is a very restricted state which is a delta function in Fourier space. In general spiral textures have complicated Fourier structure so we do not expect this state to capture the spiral phase boundaries quantitatively, only qualitatively. There is a special point in the phase diagram ($H=0$ and $A=0$) where the spiral actually has this simple form.

When $0<m_1=m_2<1$ the state is a circular cone, i.e., the magnetic moments trace out a cone in spin space as a function of $z$ when $\mb{q}=q\hat{\mb{z}}$. The circular cone found in Appendix \ref{sec.Ferromagnetic and Cone Phases} is exactly of this form with $\mb{q}=\cos\beta\hat{\mb{z}}$, $m_1=\sin\theta^*$, $\hat{\mb{e}}_1=\hat{\mb{x}}$ and $\hat{\mb{e}}_2=\hat{\mb{z}}$.

Finally we discuss the most complicated limiting case of this variational state. When $0<m_1<m_2<1$ the state (\ref{eqn.cone variation}) is an elliptic cone, i.e., the magnetic moments trace out an elliptic cone as a function of $x$, for example, when $\hat{\mb{q}}=\hat{\mb{x}}$. An elliptic cone has a base and a tip, similar to a cone, but the cross sections are ellipses rather than circles. The actual elliptic cone found using numerics is not as simple as the variational state. In particular the numerical solution does not have simple Fourier structure along the $\hat{\mb{e}}_1$ and $\hat{\mb{e}}_2$ axes, in contrast to the variational state \ref{eqn.cone variation}. However, near the polarized FM-elliptic cone and tilted FM-elliptic cone phase boundaries the elliptic cone becomes more and more like the variational state \ref{eqn.cone variation}, so the phase boundary found using this variational state is quantitatively correct.

\begin{figure}
\includegraphics[scale=0.5]{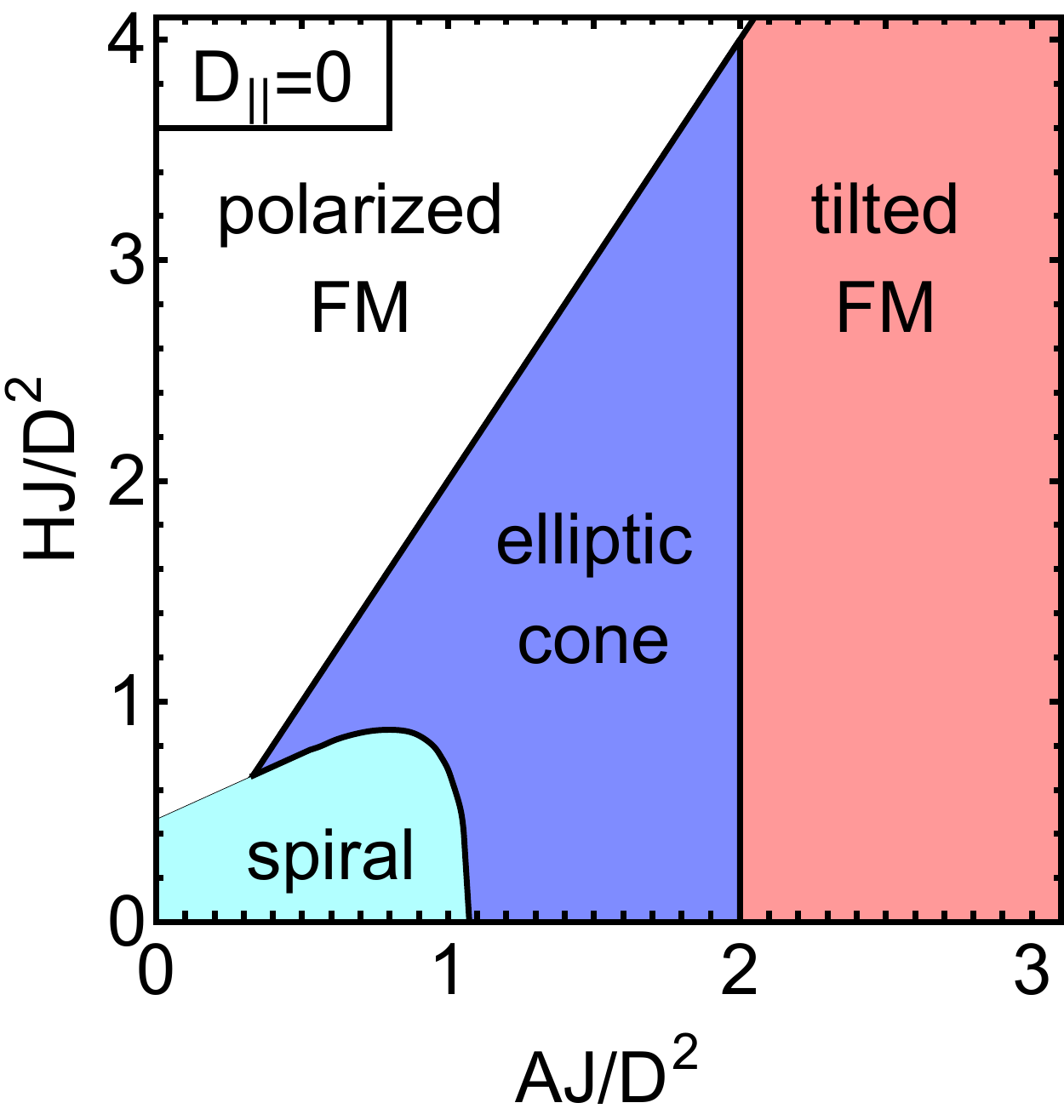}
\caption{\label{fig.phase diagram variational} Phase diagram obtained from variational Ansatz (\ref{eqn.cone variation}) for $D_\parallel=0$. Thick lines denote continuous transitions and thin lines denote first order transitions. The variational state does not allow for skyrmion phases so we do not expect them in this phase diagram. The main result is that the phase boundaries at $H=2A$ and $AJ/D^2=2$ agree to arbitrary precision with the numerical results, i.e., the polarized FM-elliptic cone and tilted FM-elliptic cone phase boundaries are identical to the boundaries in the $D_\parallel=0$ limit of Fig. 1 of the main text.}
\end{figure}

{\bf Exact integration:} It is convenient to trade the $\hat{\mb{e}}_i$ for Euler angles. This can be achieved by writing
\begin{equation}
\hat{\mb{e}}_i=\mc{R}_z(\theta_3)\mc{R}_y(\theta_2)\mc{R}_z(\theta_1)\hat{\mb{x}}_i
\end{equation}
where $\hat{\mb{x}}_1=\hat{\mb{x}}$, $\hat{\mb{x}}_2=\hat{\mb{y}}$ and $\hat{\mb{x}}_3=\hat{\mb{z}}$. This state (\ref{eqn.cone variation}) has 8 variational parameters: $\theta_1$, $\theta_2$, $\theta_3$, $m_1$, $m_2$, and $\mb{q}$. Without loss of generality we can choose $\mb{q}$ to lie in the $x$-$z$ plane leaving $7$ variational parameters. The free energy in terms of these 7 parameters is
\begin{eqnarray}\label{eqn.cone variation integrated}
F&=&F_J+F_D+F_A+F_H \\
F_J&=&\frac{1}{2}(q_x^2+q_z^2)(1-\sqrt{(1-m_1^2)(1-m_2^2)}) \nonumber \\
F_D&=&-m_1 m_2 \cos\beta(q_z\cos\theta_2+q_x\cos\theta_3\sin\theta_2) \nonumber \\
&&-m_1 m_2 \sin\beta q_x\sin\theta_2\sin\theta_3 \nonumber \\
F_A&=&\phantom{+}\frac{A}{2}\cos^2\theta_2(2-m_1^2-m_2^2)\nonumber \\
&&+\frac{A}{2}\sin^2\theta_2((m_1\cos\theta_1)^2+(m_2\sin\theta_1)^2) \nonumber \\
F_H&=&-\frac{H\cos\theta_2}{2\pi}\int_0^{2\pi}\sqrt{1-(m_1\cos u)^2-(m_2\sin u)^2}. \nonumber
\end{eqnarray}
$F_H$ can be expressed in terms of the complete elliptic integral $E(x)=\int_0^{\pi/2}\sqrt{1-x\sin^2\theta}d\theta$, i.e.,
\begin{equation}
F_H=-\frac{H}{\pi/2}\sqrt{1-m_2^2}\cos\theta_2E\left(-\frac{(m_1-m_2)(m_1+m_2)}{1-m_1^2}\right). \nonumber \\ \nonumber
\end{equation}
The free energy has been integrated analytically but the variational parameter minimization involves transcendental equations. Numerically solving these equations produces the phase diagram in Figure \ref{fig.phase diagram variational} in the $D_\parallel=0$ limit. 

Important features of this phase diagram are the polarized FM-elliptic cone and tilted FM-elliptic cone phase boundaries. These boundaries agree to arbitrary precision with the phase boundaries found using numerics, i.e., $H=2A$ for the polarized FM-elliptic cone transition and $A=2$ for the tilted FM-elliptic cone transition. The spiral phase boundary is only qualitatively correct due to the simple Fourier structure of the variational state. 

\section{Phase Transitions}\label{sec.Phase Transitions}
\begin{figure}
\includegraphics[scale=0.5]{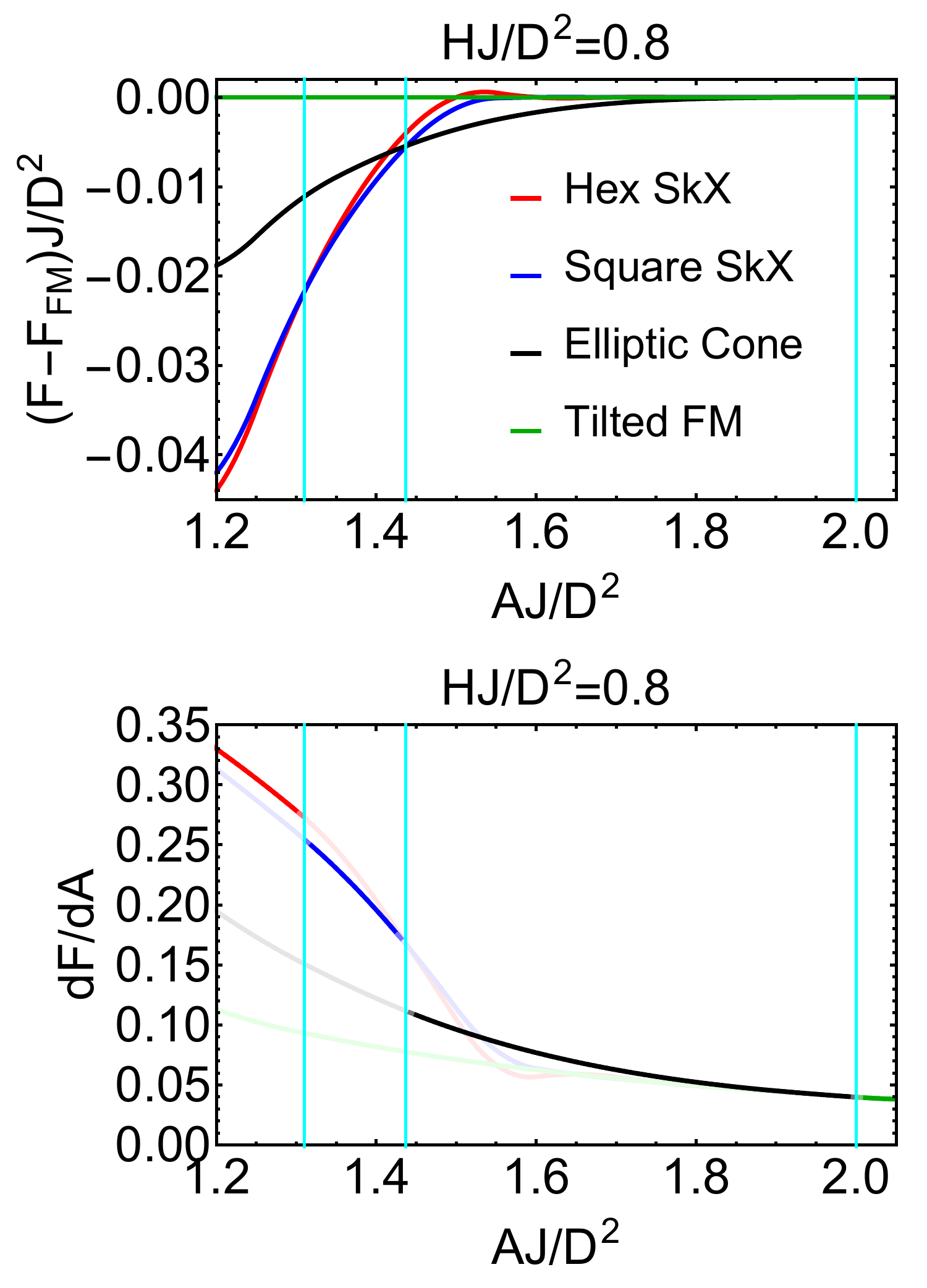}
\caption{\label{fig.phase transitions} Free energy relative to the tilted FM (top) as a function of $AJ/D^2$ at fixed $HJ/D^2=0.8$ and the derivative $dF/dA$ of the free energy (bottom) for the hexagonal skyrmion crystal (red), square skyrmion crystal (blue), elliptic cone (black) and tilted FM (green) phases. Phase transitions are marked by cyan lines. In the plot of $dF/dA$ it is easy to see which phase transitions are continuous (elliptic cone-tilted FM) and which are first order (hexagonal-square skyrmion crystal and square skyrmion crystal-elliptic cone) by examining the jump discontinuities in the derivative of the free energy. The stable phase is indicated by a darker line in the bottom figure.}
\end{figure}

In this Appendix we discuss the nature of the various phase transitions shown in Fig. 1 of the main text, where thick lines indicate continuous transitions and thin lines indicate first order transitions. All phase transitions are determined by comparing energy curves near the phase boundary (see Fig. \ref{fig.phase transitions}). Energy curves that have the same slope on both sides of the phase boundary correspond to continuous transitions, discontinuous slopes correspond to first order transitions. In the $D_\bot=0$ limit of the phase diagram we find only one continuous phase transition, between the polarized FM and the cone (vertical). In the $D_\parallel=0$ limit we find four continuous phase transitions, three of which meet at a point called a Lifshitz point~\cite{Chaikin-Lubensky}. Below we give reasoning for the continuous or first-order nature of the various transitions.

{\bf Dresselhaus limit:} The $D_\bot=0$ limit of the phase diagram has been previously studied \cite{PRBWilson2014}. 
The phase transitions are all first order except the polarized FM to cone transition, which is continuous. 
At this transition the cone radius goes continuously to zero.

The spiral-cone, spiral-skyrmion crystal and cone-skyrmion crystal transitions are all first order. In each of these transitions the phases on either side of the phase boundary have distinct broken symmetries. Generically we expect to find first order transitions between two phases with different broken symmetries.

For the first order polarized FM-spiral phase transition the polarized FM phase has no broken symmetry while the spiral phase breaks translational symmetry. Furthermore, the wavelength of the spiral remains finite at the phase boundary.

Our numerical results for the polarized FM-skyrmion crystal show that the transition is first order; however, it is an unusual first order transition with a diverging length scale associated with the optimal SkX unit cell size~\cite{PRXBanerjee2014}.

{\bf Rashba limit:} We discuss here only the phase transitions that are present in the $D_\parallel=0$ phase diagram and are not present in the $D_\bot=0$ phase diagram. 

The elliptic cone transitions continuously to the polarized FM and tilted FM phases. Near the phase boundary the radius of the cone going continuously to zero. There is a point, $A=2$ and $H=4$, where these three phases meet. This is a Lifshitz point~\cite{Chaikin-Lubensky} at which a ``symmetric'' phase, in our case
the polarized FM, meets a broken symmetry phase, the tilted FM, and a spatially modulated phase, the elliptic cone; 
see Sec.~4.6 of ref.~\onlinecite{Chaikin-Lubensky}.

The elliptic cone-spiral phase transition is also continuous, in contrast with the vertical cone-spiral phase boundary. Near the elliptic cone-spiral phase boundary the height of the cone goes continuously to zero.

The elliptic cone-hexagonal skyrmion crystal, elliptic-cone square skyrmion crystal and hexagonal-square skyrmion crystal phase boundaries are all first order with distinct broken symmetries on each side of the phase boundaries.

\section{Rashba Limit Phase Diagram}\label{sec.Rashba Limit Phase Diagram}
In appendix \ref{sec.Rashba Limit Phase Diagram} we comment in detail on how the Rashba limit phase diagram shown in the right-most
panel of Fig.~1 of the main paper, goes beyond all previous works, as mentioned in the text.

The $H=0$ results for $A>0$ can be compared by taking the $T=0$ limit of the finite temperature
analysis of ref.~\onlinecite{PRLBalents2014}. For $H=0$ and $A>0$ they find
three phases: a spiral, a cone, and an in-plane FM. These are same as the $H=0$ limit of our results
with our tilted elliptic cone simplifying to a cone whose axis is horizontal. Our phase boundaries are more accurate,
however, because, as emphasized in the text, we do not restrict attention to a single-$\mb{Q}$ spiral
as in ref.~\onlinecite{PRLBalents2014}. Further, these authors did not analyze $H \neq 0$.

For $H \neq 0$, our own earlier work~\cite{PRXBanerjee2014} presented clear evidence for the development of 
nontrivial spatial structure in the topological charge density, but in that paper we worked with a SkX
variational ansatz that forced the spins on the unit cell boundary to be pointing along the field direction.
Thus we did not have the variational freedom to see skyrmions where we must use a Chern number to
understand the quantization of topological charge. We see here
the larger hexagonal SkX region than in ref.~\onlinecite{PRXBanerjee2014} and also the square SkX.

Recently, ref.~\onlinecite{PRBBatista2015} predicted a small region of stability for
the square SkX in the $A>0$ regime, and our results are consistent with theirs as far as this 
feature of the phase diagram is concerned. However, both refs.~\onlinecite{PRXBanerjee2014,PRBBatista2015}  
missed the elliptic cone phase.

\section{Magnetic Anisotropy}\label{sec.Magnetic Anisotropy}
In appendix \ref{sec.Magnetic Anisotropy} we discuss how Rashba spin-orbit coupling gives rise to an
easy-plane magnetic anisotropy. In the paper we have a magnetic anisotropy term $\mc{F}_A=A m_z^2$ in the free energy functional, 
where $A$ is treated as a phenomenological parameter that can be 
either $A<0$ (easy-axis) or $A>0$ (easy-plane). As noted there, many mechanisms
contribute to $A$, including atomic SOC which gives rise to single-ion anisotropy and 
and dipolar interactions that lead to shape anisotropy. 

Here we comment on the anisotropy contribution of the SOC that leads to the DM terms in the
free energy functional, following refs.~\onlinecite{NaturePhysBanerjee2013,PRXBanerjee2014}.
Rashba SOC, which gives rise to a DM term 
$D_\perp \sim \lambda_{\rm soc}$ linear in the SOC coupling constant $\lambda_{\rm soc}$,
also gives rise to a compass-Kitaev anisotropy of the form
\begin{equation}\label{compass}
- A_\perp (S^x_i S^x_{i+y} + S^y_i S^y_{i+x}).
\end{equation}
This term is often ignored in the literature because $A_\perp \sim \lambda_{\rm soc}^2$ is small.
However, this argument is flawed because the energetic contribution of the DM term is of
order $D_\perp^2/J$ and hence of the same order as that of $A_\perp$.
In fact, one can show that for a large class of exchange mechanisms
$A_\perp J/D_\perp^2 \simeq 1/2$ in the limit of weak SOC; see ref.~\onlinecite{PRXBanerjee2014}
and references therein. In the continuum limit, and ignoring higher order derivative terms,
eq.~(\ref{compass}) leads to $+|A_\perp| m_z^2$. This is an
important easy-plane contribution to the total anisotropy.

What about the analogous term arising from Dresselhaus SOC?
This is of the form
\begin{equation}
- A_\parallel (S^x_i S^x_{i+x} + S^y_i S^y_{i+y} + S^z_i S^z_{i+z})
\end{equation}
with $A_\parallel J/D_\parallel^2 = 1/2$. 
If we take the continuum limit, retain just the order $m^2$ terms
and ignore higher order derivative terms, we get $A_\parallel (m_x^2 + m_y^2 + m_z^2)$
which is just an additive constant of no consequence, since $\mb{m}^2 = 1$.
So it is only the Rashba SOC that gives rise to the interesting easy-plane anisotropy,
as one can see from symmetry alone.

\end{appendix}


\begin{thebibliography}{32}%
\makeatletter
\providecommand \@ifxundefined [1]{%
 \@ifx{#1\undefined}
}%
\providecommand \@ifnum [1]{%
 \ifnum #1\expandafter \@firstoftwo
 \else \expandafter \@secondoftwo
 \fi
}%
\providecommand \@ifx [1]{%
 \ifx #1\expandafter \@firstoftwo
 \else \expandafter \@secondoftwo
 \fi
}%
\providecommand \natexlab [1]{#1}%
\providecommand \enquote  [1]{``#1''}%
\providecommand \bibnamefont  [1]{#1}%
\providecommand \bibfnamefont [1]{#1}%
\providecommand \citenamefont [1]{#1}%
\providecommand \href@noop [0]{\@secondoftwo}%
\providecommand \href [0]{\begingroup \@sanitize@url \@href}%
\providecommand \@href[1]{\@@startlink{#1}\@@href}%
\providecommand \@@href[1]{\endgroup#1\@@endlink}%
\providecommand \@sanitize@url [0]{\catcode `\\12\catcode `\$12\catcode
  `\&12\catcode `\#12\catcode `\^12\catcode `\_12\catcode `\%12\relax}%
\providecommand \@@startlink[1]{}%
\providecommand \@@endlink[0]{}%
\providecommand \url  [0]{\begingroup\@sanitize@url \@url }%
\providecommand \@url [1]{\endgroup\@href {#1}{\urlprefix }}%
\providecommand \urlprefix  [0]{URL }%
\providecommand \Eprint [0]{\href }%
\providecommand \doibase [0]{http://dx.doi.org/}%
\providecommand \selectlanguage [0]{\@gobble}%
\providecommand \bibinfo  [0]{\@secondoftwo}%
\providecommand \bibfield  [0]{\@secondoftwo}%
\providecommand \translation [1]{[#1]}%
\providecommand \BibitemOpen [0]{}%
\providecommand \bibitemStop [0]{}%
\providecommand \bibitemNoStop [0]{.\EOS\space}%
\providecommand \EOS [0]{\spacefactor3000\relax}%
\providecommand \BibitemShut  [1]{\csname bibitem#1\endcsname}%
\let\auto@bib@innerbib\@empty
\bibitem [{\citenamefont {Bogdanov}\ and\ \citenamefont
  {Hubert}(1994)}]{JMMMBogdanov1994}%
  \BibitemOpen
  \bibfield  {author} {\bibinfo {author} {\bibfnamefont {A.}~\bibnamefont
  {Bogdanov}}\ and\ \bibinfo {author} {\bibfnamefont {A.}~\bibnamefont
  {Hubert}},\ }\href {\doibase http://dx.doi.org/10.1016/0304-8853(94)90046-9}
  {\bibfield  {journal} {\bibinfo  {journal} {J. Magn. Magn. Mat.}\ }\textbf
  {\bibinfo {volume} {138}},\ \bibinfo {pages} {255 } (\bibinfo {year}
  {1994})}\BibitemShut {NoStop}%
\bibitem [{\citenamefont {R{\"o}{\ss}ler}\ \emph {et~al.}(2006)\citenamefont
  {R{\"o}{\ss}ler}, \citenamefont {Bogdanov},\ and\ \citenamefont
  {Pfleiderer}}]{NatureRossler2006}%
  \BibitemOpen
  \bibfield  {author} {\bibinfo {author} {\bibfnamefont {U.}~\bibnamefont
  {R{\"o}{\ss}ler}}, \bibinfo {author} {\bibfnamefont {A.}~\bibnamefont
  {Bogdanov}}, \ and\ \bibinfo {author} {\bibfnamefont {C.}~\bibnamefont
  {Pfleiderer}},\ }\href@noop {} {\bibfield  {journal} {\bibinfo  {journal}
  {Nature}\ }\textbf {\bibinfo {volume} {442}},\ \bibinfo {pages} {797}
  (\bibinfo {year} {2006})}\BibitemShut {NoStop}%
\bibitem [{\citenamefont {Nagaosa}\ and\ \citenamefont
  {Tokura}(2013)}]{NatureNanoNagaosa2013}%
  \BibitemOpen
  \bibfield  {author} {\bibinfo {author} {\bibfnamefont {N.}~\bibnamefont
  {Nagaosa}}\ and\ \bibinfo {author} {\bibfnamefont {Y.}~\bibnamefont
  {Tokura}},\ }\href@noop {} {\bibfield  {journal} {\bibinfo  {journal} {Nature
  Nanotech.}\ }\textbf {\bibinfo {volume} {8}},\ \bibinfo {pages} {899}
  (\bibinfo {year} {2013})}\BibitemShut {NoStop}%
\bibitem [{\citenamefont {Neubauer}\ \emph {et~al.}(2009)\citenamefont
  {Neubauer}, \citenamefont {Pfleiderer}, \citenamefont {Binz}, \citenamefont
  {Rosch}, \citenamefont {Ritz}, \citenamefont {Niklowitz},\ and\ \citenamefont
  {B\"oni}}]{PRLNeubauer2009}%
  \BibitemOpen
  \bibfield  {author} {\bibinfo {author} {\bibfnamefont {A.}~\bibnamefont
  {Neubauer}}, \bibinfo {author} {\bibfnamefont {C.}~\bibnamefont
  {Pfleiderer}}, \bibinfo {author} {\bibfnamefont {B.}~\bibnamefont {Binz}},
  \bibinfo {author} {\bibfnamefont {A.}~\bibnamefont {Rosch}}, \bibinfo
  {author} {\bibfnamefont {R.}~\bibnamefont {Ritz}}, \bibinfo {author}
  {\bibfnamefont {P.~G.}\ \bibnamefont {Niklowitz}}, \ and\ \bibinfo {author}
  {\bibfnamefont {P.}~\bibnamefont {B\"oni}},\ }\href {\doibase
  10.1103/PhysRevLett.102.186602} {\bibfield  {journal} {\bibinfo  {journal}
  {Phys. Rev. Lett.}\ }\textbf {\bibinfo {volume} {102}},\ \bibinfo {pages}
  {186602} (\bibinfo {year} {2009})}\BibitemShut {NoStop}%
\bibitem [{\citenamefont {Lee}\ \emph {et~al.}(2009)\citenamefont {Lee},
  \citenamefont {Kang}, \citenamefont {Onose}, \citenamefont {Tokura},\ and\
  \citenamefont {Ong}}]{PRLLee2009}%
  \BibitemOpen
  \bibfield  {author} {\bibinfo {author} {\bibfnamefont {M.}~\bibnamefont
  {Lee}}, \bibinfo {author} {\bibfnamefont {W.}~\bibnamefont {Kang}}, \bibinfo
  {author} {\bibfnamefont {Y.}~\bibnamefont {Onose}}, \bibinfo {author}
  {\bibfnamefont {Y.}~\bibnamefont {Tokura}}, \ and\ \bibinfo {author}
  {\bibfnamefont {N.}~\bibnamefont {Ong}},\ }\href {\doibase
  10.1103/PhysRevLett.102.186601} {\bibfield  {journal} {\bibinfo  {journal}
  {Phys. Rev. Lett.}\ }\textbf {\bibinfo {volume} {102}},\ \bibinfo {pages}
  {186601} (\bibinfo {year} {2009})}\BibitemShut {NoStop}%
\bibitem [{\citenamefont {Ritz}\ \emph {et~al.}(2013)\citenamefont {Ritz},
  \citenamefont {Halder}, \citenamefont {Wagner}, \citenamefont {Franz},
  \citenamefont {Bauer},\ and\ \citenamefont {Pfleiderer}}]{NatureRitz2013}%
  \BibitemOpen
  \bibfield  {author} {\bibinfo {author} {\bibfnamefont {R.}~\bibnamefont
  {Ritz}}, \bibinfo {author} {\bibfnamefont {M.}~\bibnamefont {Halder}},
  \bibinfo {author} {\bibfnamefont {M.}~\bibnamefont {Wagner}}, \bibinfo
  {author} {\bibfnamefont {C.}~\bibnamefont {Franz}}, \bibinfo {author}
  {\bibfnamefont {A.}~\bibnamefont {Bauer}}, \ and\ \bibinfo {author}
  {\bibfnamefont {C.}~\bibnamefont {Pfleiderer}},\ }\href@noop {} {\bibfield
  {journal} {\bibinfo  {journal} {Nature}\ }\textbf {\bibinfo {volume} {497}},\
  \bibinfo {pages} {231} (\bibinfo {year} {2013})}\BibitemShut {NoStop}%
\bibitem [{\citenamefont {Romming}\ \emph {et~al.}(2013)\citenamefont
  {Romming}, \citenamefont {Hanneken}, \citenamefont {Menzel}, \citenamefont
  {Bickel}, \citenamefont {Wolter}, \citenamefont {von Bergmann}, \citenamefont
  {Kubetzka},\ and\ \citenamefont {Wiesendanger}}]{ScienceRomming2013}%
  \BibitemOpen
  \bibfield  {author} {\bibinfo {author} {\bibfnamefont {N.}~\bibnamefont
  {Romming}}, \bibinfo {author} {\bibfnamefont {C.}~\bibnamefont {Hanneken}},
  \bibinfo {author} {\bibfnamefont {M.}~\bibnamefont {Menzel}}, \bibinfo
  {author} {\bibfnamefont {J.}~\bibnamefont {Bickel}}, \bibinfo {author}
  {\bibfnamefont {B.}~\bibnamefont {Wolter}}, \bibinfo {author} {\bibfnamefont
  {K.}~\bibnamefont {von Bergmann}}, \bibinfo {author} {\bibfnamefont
  {A.}~\bibnamefont {Kubetzka}}, \ and\ \bibinfo {author} {\bibfnamefont
  {R.}~\bibnamefont {Wiesendanger}},\ }\href@noop {} {\bibfield  {journal}
  {\bibinfo  {journal} {Science}\ }\textbf {\bibinfo {volume} {341}},\ \bibinfo
  {pages} {636} (\bibinfo {year} {2013})}\BibitemShut {NoStop}%
\bibitem [{\citenamefont {Jonietz}\ \emph {et~al.}(2010)\citenamefont
  {Jonietz}, \citenamefont {M{\"u}hlbauer}, \citenamefont {Pfleiderer},
  \citenamefont {Neubauer}, \citenamefont {M{\"u}nzer}, \citenamefont {Bauer},
  \citenamefont {Adams}, \citenamefont {Georgii}, \citenamefont {B{\"o}ni},
  \citenamefont {Duine}, \citenamefont {Everschor}, \citenamefont {Garst},\
  and\ \citenamefont {Rosch}}]{ScienceJonietz2010}%
  \BibitemOpen
  \bibfield  {author} {\bibinfo {author} {\bibfnamefont {F.}~\bibnamefont
  {Jonietz}}, \bibinfo {author} {\bibfnamefont {S.}~\bibnamefont
  {M{\"u}hlbauer}}, \bibinfo {author} {\bibfnamefont {C.}~\bibnamefont
  {Pfleiderer}}, \bibinfo {author} {\bibfnamefont {A.}~\bibnamefont
  {Neubauer}}, \bibinfo {author} {\bibfnamefont {W.}~\bibnamefont
  {M{\"u}nzer}}, \bibinfo {author} {\bibfnamefont {A.}~\bibnamefont {Bauer}},
  \bibinfo {author} {\bibfnamefont {T.}~\bibnamefont {Adams}}, \bibinfo
  {author} {\bibfnamefont {R.}~\bibnamefont {Georgii}}, \bibinfo {author}
  {\bibfnamefont {P.}~\bibnamefont {B{\"o}ni}}, \bibinfo {author}
  {\bibfnamefont {R.}~\bibnamefont {Duine}}, \bibinfo {author} {\bibfnamefont
  {K.}~\bibnamefont {Everschor}}, \bibinfo {author} {\bibfnamefont
  {M.}~\bibnamefont {Garst}}, \ and\ \bibinfo {author} {\bibfnamefont
  {A.}~\bibnamefont {Rosch}},\ }\href@noop {} {\bibfield  {journal} {\bibinfo
  {journal} {Science}\ }\textbf {\bibinfo {volume} {330}},\ \bibinfo {pages}
  {1648} (\bibinfo {year} {2010})}\BibitemShut {NoStop}%
\bibitem [{\citenamefont {Mühlbauer}\ \emph {et~al.}(2009)\citenamefont
  {Mühlbauer}, \citenamefont {Binz}, \citenamefont {Jonietz}, \citenamefont
  {Pfleiderer}, \citenamefont {Rosch}, \citenamefont {Neubauer}, \citenamefont
  {Georgii},\ and\ \citenamefont {Böni}}]{ScienceMuhlbauer2009}%
  \BibitemOpen
  \bibfield  {author} {\bibinfo {author} {\bibfnamefont {S.}~\bibnamefont
  {M\"{u}hlbauer}}, \bibinfo {author} {\bibfnamefont {B.}~\bibnamefont {Binz}},
  \bibinfo {author} {\bibfnamefont {F.}~\bibnamefont {Jonietz}}, \bibinfo
  {author} {\bibfnamefont {C.}~\bibnamefont {Pfleiderer}}, \bibinfo {author}
  {\bibfnamefont {A.}~\bibnamefont {Rosch}}, \bibinfo {author} {\bibfnamefont
  {A.}~\bibnamefont {Neubauer}}, \bibinfo {author} {\bibfnamefont
  {R.}~\bibnamefont {Georgii}}, \ and\ \bibinfo {author} {\bibfnamefont
  {P.}~\bibnamefont {B\"{o}ni}},\ }\href {\doibase 10.1126/science.1166767}
  {\bibfield  {journal} {\bibinfo  {journal} {Science}\ }\textbf {\bibinfo
  {volume} {323}},\ \bibinfo {pages} {915} (\bibinfo {year}
  {2009})}\BibitemShut {NoStop}%
\bibitem [{\citenamefont {Yu}\ \emph {et~al.}(2011)\citenamefont {Yu},
  \citenamefont {Kanazawa}, \citenamefont {Onose}, \citenamefont {Kimoto},
  \citenamefont {Zhang}, \citenamefont {Ishiwata}, \citenamefont {Matsui},\
  and\ \citenamefont {Tokura}}]{NatureMatYu2011}%
  \BibitemOpen
  \bibfield  {author} {\bibinfo {author} {\bibfnamefont {X.}~\bibnamefont
  {Yu}}, \bibinfo {author} {\bibfnamefont {N.}~\bibnamefont {Kanazawa}},
  \bibinfo {author} {\bibfnamefont {Y.}~\bibnamefont {Onose}}, \bibinfo
  {author} {\bibfnamefont {K.}~\bibnamefont {Kimoto}}, \bibinfo {author}
  {\bibfnamefont {W.}~\bibnamefont {Zhang}}, \bibinfo {author} {\bibfnamefont
  {S.}~\bibnamefont {Ishiwata}}, \bibinfo {author} {\bibfnamefont
  {Y.}~\bibnamefont {Matsui}}, \ and\ \bibinfo {author} {\bibfnamefont
  {Y.}~\bibnamefont {Tokura}},\ }\href@noop {} {\bibfield  {journal} {\bibinfo
  {journal} {Nature Mater.}\ }\textbf {\bibinfo {volume} {10}},\ \bibinfo
  {pages} {106} (\bibinfo {year} {2011})}\BibitemShut {NoStop}%
\bibitem [{\citenamefont {Seki}\ \emph {et~al.}(2012)\citenamefont {Seki},
  \citenamefont {Yu}, \citenamefont {Ishiwata},\ and\ \citenamefont
  {Tokura}}]{ScienceSeki2012}%
  \BibitemOpen
  \bibfield  {author} {\bibinfo {author} {\bibfnamefont {S.}~\bibnamefont
  {Seki}}, \bibinfo {author} {\bibfnamefont {X.}~\bibnamefont {Yu}}, \bibinfo
  {author} {\bibfnamefont {S.}~\bibnamefont {Ishiwata}}, \ and\ \bibinfo
  {author} {\bibfnamefont {Y.}~\bibnamefont {Tokura}},\ }\href {\doibase
  10.1126/science.1214143} {\bibfield  {journal} {\bibinfo  {journal}
  {Science}\ }\textbf {\bibinfo {volume} {336}},\ \bibinfo {pages} {198}
  (\bibinfo {year} {2012})}\BibitemShut {NoStop}%
\bibitem{Binz2006}
B.~Binz, A.~Vishwanath and V.~Aji, Phys.~Rev.~Lett.~{\bf 96}, 207202 (2006).

\bibitem [{\citenamefont {Wilson}\ \emph {et~al.}(2014)\citenamefont {Wilson},
  \citenamefont {Butenko}, \citenamefont {Bogdanov},\ and\ \citenamefont
  {Monchesky}}]{PRBWilson2014}%
  \BibitemOpen
  \bibfield  {author} {\bibinfo {author} {\bibfnamefont {M.}~\bibnamefont
  {Wilson}}, \bibinfo {author} {\bibfnamefont {A.}~\bibnamefont {Butenko}},
  \bibinfo {author} {\bibfnamefont {A.}~\bibnamefont {Bogdanov}}, \ and\
  \bibinfo {author} {\bibfnamefont {T.}~\bibnamefont {Monchesky}},\ }\href@noop
  {} {\bibfield  {journal} {\bibinfo  {journal} {Phys. Rev. B}\ }\textbf
  {\bibinfo {volume} {89}},\ \bibinfo {pages} {094411} (\bibinfo {year}
  {2014})}\BibitemShut {NoStop}%
\bibitem [{\citenamefont {Huang}\ and\ \citenamefont
  {Chien}(2012)}]{PRLHuang2012}%
  \BibitemOpen
  \bibfield  {author} {\bibinfo {author} {\bibfnamefont {S.~X.}\ \bibnamefont
  {Huang}}\ and\ \bibinfo {author} {\bibfnamefont {C.~L.}\ \bibnamefont
  {Chien}},\ }\href {\doibase 10.1103/PhysRevLett.108.267201} {\bibfield
  {journal} {\bibinfo  {journal} {Phys. Rev. Lett.}\ }\textbf {\bibinfo
  {volume} {108}},\ \bibinfo {pages} {267201} (\bibinfo {year}
  {2012})}\BibitemShut {NoStop}%
\bibitem [{\citenamefont {Tonomura}\ \emph {et~al.}(2012)\citenamefont
  {Tonomura}, \citenamefont {Yu}, \citenamefont {Yanagisawa}, \citenamefont
  {Matsuda}, \citenamefont {Onose}, \citenamefont {Kanazawa}, \citenamefont
  {Park},\ and\ \citenamefont {Tokura}}]{NanoLettTonomura2012}%
  \BibitemOpen
  \bibfield  {author} {\bibinfo {author} {\bibfnamefont {A.}~\bibnamefont
  {Tonomura}}, \bibinfo {author} {\bibfnamefont {X.}~\bibnamefont {Yu}},
  \bibinfo {author} {\bibfnamefont {K.}~\bibnamefont {Yanagisawa}}, \bibinfo
  {author} {\bibfnamefont {T.}~\bibnamefont {Matsuda}}, \bibinfo {author}
  {\bibfnamefont {Y.}~\bibnamefont {Onose}}, \bibinfo {author} {\bibfnamefont
  {N.}~\bibnamefont {Kanazawa}}, \bibinfo {author} {\bibfnamefont {H.~S.}\
  \bibnamefont {Park}}, \ and\ \bibinfo {author} {\bibfnamefont
  {Y.}~\bibnamefont {Tokura}},\ }\href {\doibase 10.1021/nl300073m} {\bibfield
  {journal} {\bibinfo  {journal} {Nano letters}\ }\textbf {\bibinfo {volume}
  {12}},\ \bibinfo {pages} {1673} (\bibinfo {year} {2012})}\BibitemShut
  {NoStop}%
\bibitem [{\citenamefont {Yu}\ \emph {et~al.}(2010)\citenamefont {Yu},
  \citenamefont {Onose}, \citenamefont {Kanazawa}, \citenamefont {Park},
  \citenamefont {Han}, \citenamefont {Matsui}, \citenamefont {Nagaosa},\ and\
  \citenamefont {Tokura}}]{NatYu2010}%
  \BibitemOpen
  \bibfield  {author} {\bibinfo {author} {\bibfnamefont {X.~Z.}\ \bibnamefont
  {Yu}}, \bibinfo {author} {\bibfnamefont {Y.}~\bibnamefont {Onose}}, \bibinfo
  {author} {\bibfnamefont {N.}~\bibnamefont {Kanazawa}}, \bibinfo {author}
  {\bibfnamefont {J.~H.}\ \bibnamefont {Park}}, \bibinfo {author}
  {\bibfnamefont {J.~H.}\ \bibnamefont {Han}}, \bibinfo {author} {\bibfnamefont
  {Y.}~\bibnamefont {Matsui}}, \bibinfo {author} {\bibfnamefont
  {N.}~\bibnamefont {Nagaosa}}, \ and\ \bibinfo {author} {\bibfnamefont
  {Y.}~\bibnamefont {Tokura}},\ }\href {\doibase 10.1038/nature09124}
  {\bibfield  {journal} {\bibinfo  {journal} {Nature}\ }\textbf {\bibinfo
  {volume} {465}},\ \bibinfo {pages} {901} (\bibinfo {year}
  {2010})}\BibitemShut {NoStop}%
\bibitem [{\citenamefont {Heinze}\ \emph {et~al.}(2011)\citenamefont {Heinze},
  \citenamefont {von Bergmann}, \citenamefont {Menzel}, \citenamefont {Brede},
  \citenamefont {Kubetzka}, \citenamefont {Wiesendanger}, \citenamefont
  {Bihlmayer},\ and\ \citenamefont {Bl{\"u}gel}}]{NatPhysHeinze2011}%
  \BibitemOpen
  \bibfield  {author} {\bibinfo {author} {\bibfnamefont {S.}~\bibnamefont
  {Heinze}}, \bibinfo {author} {\bibfnamefont {K.}~\bibnamefont {von
  Bergmann}}, \bibinfo {author} {\bibfnamefont {M.}~\bibnamefont {Menzel}},
  \bibinfo {author} {\bibfnamefont {J.}~\bibnamefont {Brede}}, \bibinfo
  {author} {\bibfnamefont {A.}~\bibnamefont {Kubetzka}}, \bibinfo {author}
  {\bibfnamefont {R.}~\bibnamefont {Wiesendanger}}, \bibinfo {author}
  {\bibfnamefont {G.}~\bibnamefont {Bihlmayer}}, \ and\ \bibinfo {author}
  {\bibfnamefont {S.}~\bibnamefont {Bl{\"u}gel}},\ }\href {\doibase
  10.1038/nphys2045} {\bibfield  {journal} {\bibinfo  {journal} {Nature
  Physics}\ }\textbf {\bibinfo {volume} {7}},\ \bibinfo {pages} {713} (\bibinfo
  {year} {2011})}\BibitemShut {NoStop}%
\bibitem [{\citenamefont {Banerjee}\ \emph {et~al.}(2013)\citenamefont
  {Banerjee}, \citenamefont {Erten},\ and\ \citenamefont
  {Randeria}}]{NaturePhysBanerjee2013}%
  \BibitemOpen
  \bibfield  {author} {\bibinfo {author} {\bibfnamefont {S.}~\bibnamefont
  {Banerjee}}, \bibinfo {author} {\bibfnamefont {O.}~\bibnamefont {Erten}}, \
  and\ \bibinfo {author} {\bibfnamefont {M.}~\bibnamefont {Randeria}},\
  }\href@noop {} {\bibfield  {journal} {\bibinfo  {journal} {Nature Phys.}\
  }\textbf {\bibinfo {volume} {9}},\ \bibinfo {pages} {626} (\bibinfo {year}
  {2013})}\BibitemShut {NoStop}%
\bibitem [{\citenamefont {Banerjee}\ \emph {et~al.}(2014)\citenamefont
  {Banerjee}, \citenamefont {Rowland}, \citenamefont {Erten},\ and\
  \citenamefont {Randeria}}]{PRXBanerjee2014}%
  \BibitemOpen
  \bibfield  {author} {\bibinfo {author} {\bibfnamefont {S.}~\bibnamefont
  {Banerjee}}, \bibinfo {author} {\bibfnamefont {J.}~\bibnamefont {Rowland}},
  \bibinfo {author} {\bibfnamefont {O.}~\bibnamefont {Erten}}, \ and\ \bibinfo
  {author} {\bibfnamefont {M.}~\bibnamefont {Randeria}},\ }\href {\doibase
  10.1103/PhysRevX.4.031045} {\bibfield  {journal} {\bibinfo  {journal} {Phys.
  Rev. X}\ }\textbf {\bibinfo {volume} {4}},\ \bibinfo {pages} {031045}
  (\bibinfo {year} {2014})}\BibitemShut {NoStop}%
\bibitem [{\citenamefont {Li}\ \emph {et~al.}(2014)\citenamefont {Li},
  \citenamefont {Liu},\ and\ \citenamefont {Balents}}]{PRLBalents2014}%
  \BibitemOpen
  \bibfield  {author} {\bibinfo {author} {\bibfnamefont {X.}~\bibnamefont
  {Li}}, \bibinfo {author} {\bibfnamefont {W.}~\bibnamefont {Liu}}, \ and\
  \bibinfo {author} {\bibfnamefont {L.}~\bibnamefont {Balents}},\ }\href@noop
  {} {\bibfield  {journal} {\bibinfo  {journal} {Phys. Rev. Lett.}\ }\textbf
  {\bibinfo {volume} {112}},\ \bibinfo {pages} {067202} (\bibinfo {year}
  {2014})}\BibitemShut {NoStop}%
\bibitem [{\citenamefont {Yokoyama}\ and\ \citenamefont
  {Linder}(2015)}]{Yokoyama2015}%
  \BibitemOpen
  \bibfield  {author} {\bibinfo {author} {\bibfnamefont {T.}~\bibnamefont
  {Yokoyama}}\ and\ \bibinfo {author} {\bibfnamefont {J.}~\bibnamefont
  {Linder}},\ }\href {\doibase 10.1103/PhysRevB.92.060503} {\bibfield
  {journal} {\bibinfo  {journal} {Phys. Rev. B}\ }\textbf {\bibinfo {volume}
  {92}},\ \bibinfo {pages} {060503} (\bibinfo {year} {2015})}\BibitemShut
  {NoStop}%
\bibitem [{Pri()}]{PrivateCommunicationBatista}%
  \BibitemOpen
  \href@noop {} {}\bibinfo {note} {C. Batista private
  communication}\BibitemShut {NoStop}%
\bibitem [{\citenamefont {Spaldin}\ \emph {et~al.}(2008)\citenamefont
  {Spaldin}, \citenamefont {Fiebig},\ and\ \citenamefont
  {Mostovoy}}]{JPhysCondMatSpaldin2015}%
  \BibitemOpen
  \bibfield  {author} {\bibinfo {author} {\bibfnamefont {N.}~\bibnamefont
  {Spaldin}}, \bibinfo {author} {\bibfnamefont {M.}~\bibnamefont {Fiebig}}, \
  and\ \bibinfo {author} {\bibfnamefont {M.}~\bibnamefont {Mostovoy}},\ }\href
  {http://stacks.iop.org/0953-8984/20/i=43/a=434203} {\bibfield  {journal}
  {\bibinfo  {journal} {Journal of Physics: Condensed Matter}\ }\textbf
  {\bibinfo {volume} {20}},\ \bibinfo {pages} {434203} (\bibinfo {year}
  {2008})}\BibitemShut {NoStop}%
\bibitem{Castan2012}
T.~Castan, A.~Planes, and A.~Saxena, Phys.~Rev.~B {\bf 85}, 144429 (2012).
\bibitem [{\citenamefont {Lin}\ \emph {et~al.}(2015)\citenamefont {Lin},
  \citenamefont {Saxena},\ and\ \citenamefont {Batista}}]{PRBBatista2015}%
  \BibitemOpen
  \bibfield  {author} {\bibinfo {author} {\bibfnamefont {S.~Z.}\ \bibnamefont
  {Lin}}, \bibinfo {author} {\bibfnamefont {A.}~\bibnamefont {Saxena}}, \ and\
  \bibinfo {author} {\bibfnamefont {C.~D.}\ \bibnamefont {Batista}},\ }\href
  {\doibase 10.1103/PhysRevB.91.224407} {\bibfield  {journal} {\bibinfo
  {journal} {Phys. Rev. B}\ }\textbf {\bibinfo {volume} {91}},\ \bibinfo
  {pages} {224407} (\bibinfo {year} {2015})}\BibitemShut {NoStop}%
\bibitem [{\citenamefont {Chaikin}\ and\ \citenamefont
  {Lubensky}(1995)}]{Chaikin-Lubensky}%
  \BibitemOpen
  \bibfield  {author} {\bibinfo {author} {\bibfnamefont {P.~M.}\ \bibnamefont
  {Chaikin}}\ and\ \bibinfo {author} {\bibfnamefont {T.~C.}\ \bibnamefont
  {Lubensky}},\ }\href@noop {} {\emph {\bibinfo {title} {Principles of
  Condensed Matter Physics}}}\ (\bibinfo  {publisher} {Cambridge University
  Press},\ \bibinfo {year} {1995})\BibitemShut {NoStop}%
\bibitem [{\citenamefont {Wilson}\ \emph {et~al.}(2012)\citenamefont {Wilson},
  \citenamefont {Karhu}, \citenamefont {Quigley}, \citenamefont {R\"o\ss{}ler},
  \citenamefont {Butenko}, \citenamefont {Bogdanov}, \citenamefont
  {Robertson},\ and\ \citenamefont {Monchesky}}]{Wilson2012}%
  \BibitemOpen
  \bibfield  {author} {\bibinfo {author} {\bibfnamefont {M.~N.}\ \bibnamefont
  {Wilson}}, \bibinfo {author} {\bibfnamefont {E.~A.}\ \bibnamefont {Karhu}},
  \bibinfo {author} {\bibfnamefont {A.~S.}\ \bibnamefont {Quigley}}, \bibinfo
  {author} {\bibfnamefont {U.~K.}\ \bibnamefont {R\"o\ss{}ler}}, \bibinfo
  {author} {\bibfnamefont {A.~B.}\ \bibnamefont {Butenko}}, \bibinfo {author}
  {\bibfnamefont {A.~N.}\ \bibnamefont {Bogdanov}}, \bibinfo {author}
  {\bibfnamefont {M.~D.}\ \bibnamefont {Robertson}}, \ and\ \bibinfo {author}
  {\bibfnamefont {T.~L.}\ \bibnamefont {Monchesky}},\ }\href {\doibase
  10.1103/PhysRevB.86.144420} {\bibfield  {journal} {\bibinfo  {journal} {Phys.
  Rev. B}\ }\textbf {\bibinfo {volume} {86}},\ \bibinfo {pages} {144420}
  (\bibinfo {year} {2012})}\BibitemShut {NoStop}%
\bibitem [{\citenamefont {Karhu}\ \emph {et~al.}(2012)\citenamefont {Karhu},
  \citenamefont {R\"o\ss{}ler}, \citenamefont {Bogdanov}, \citenamefont
  {Kahwaji}, \citenamefont {Kirby}, \citenamefont {Fritzsche}, \citenamefont
  {Robertson}, \citenamefont {Majkrzak},\ and\ \citenamefont
  {Monchesky}}]{Karhu2012}%
  \BibitemOpen
  \bibfield  {author} {\bibinfo {author} {\bibfnamefont {E.~A.}\ \bibnamefont
  {Karhu}}, \bibinfo {author} {\bibfnamefont {U.~K.}\ \bibnamefont
  {R\"o\ss{}ler}}, \bibinfo {author} {\bibfnamefont {A.~N.}\ \bibnamefont
  {Bogdanov}}, \bibinfo {author} {\bibfnamefont {S.}~\bibnamefont {Kahwaji}},
  \bibinfo {author} {\bibfnamefont {B.~J.}\ \bibnamefont {Kirby}}, \bibinfo
  {author} {\bibfnamefont {H.}~\bibnamefont {Fritzsche}}, \bibinfo {author}
  {\bibfnamefont {M.~D.}\ \bibnamefont {Robertson}}, \bibinfo {author}
  {\bibfnamefont {C.~F.}\ \bibnamefont {Majkrzak}}, \ and\ \bibinfo {author}
  {\bibfnamefont {T.~L.}\ \bibnamefont {Monchesky}},\ }\href {\doibase
  10.1103/PhysRevB.85.094429} {\bibfield  {journal} {\bibinfo  {journal} {Phys.
  Rev. B}\ }\textbf {\bibinfo {volume} {85}},\ \bibinfo {pages} {094429}
  (\bibinfo {year} {2012})}\BibitemShut {NoStop}%
\bibitem [{\citenamefont {Rybakov}\ \emph {et~al.}(2013)\citenamefont
  {Rybakov}, \citenamefont {Borisov},\ and\ \citenamefont
  {Bogdanov}}]{Rybakov2013}%
  \BibitemOpen
  \bibfield  {author} {\bibinfo {author} {\bibfnamefont {F.~N.}\ \bibnamefont
  {Rybakov}}, \bibinfo {author} {\bibfnamefont {A.~B.}\ \bibnamefont
  {Borisov}}, \ and\ \bibinfo {author} {\bibfnamefont {A.~N.}\ \bibnamefont
  {Bogdanov}},\ }\href {\doibase 10.1103/PhysRevB.87.094424} {\bibfield
  {journal} {\bibinfo  {journal} {Phys. Rev. B}\ }\textbf {\bibinfo {volume}
  {87}},\ \bibinfo {pages} {094424} (\bibinfo {year} {2013})}\BibitemShut
  {NoStop}%
\bibitem [{\citenamefont {Rybakov}\ \emph {et~al.}(2015)\citenamefont
  {Rybakov}, \citenamefont {Borisov}, \citenamefont {Bl\"ugel},\ and\
  \citenamefont {Kiselev}}]{Rybakov2015}%
  \BibitemOpen
  \bibfield  {author} {\bibinfo {author} {\bibfnamefont {F.~N.}\ \bibnamefont
  {Rybakov}}, \bibinfo {author} {\bibfnamefont {A.~B.}\ \bibnamefont
  {Borisov}}, \bibinfo {author} {\bibfnamefont {S.}~\bibnamefont {Bl\"ugel}}, \
  and\ \bibinfo {author} {\bibfnamefont {N.~S.}\ \bibnamefont {Kiselev}},\
  }\href {\doibase 10.1103/PhysRevLett.115.117201} {\bibfield  {journal}
  {\bibinfo  {journal} {Phys. Rev. Lett.}\ }\textbf {\bibinfo {volume} {115}},\
  \bibinfo {pages} {117201} (\bibinfo {year} {2015})}\BibitemShut {NoStop}%
\bibitem [{\citenamefont {Ishizaka}\ \emph {et~al.}(2011)\citenamefont
  {Ishizaka}, \citenamefont {Bahramy}, \citenamefont {Murakawa}, \citenamefont
  {Sakano}, \citenamefont {Shimojima}, \citenamefont {Sonobe}, \citenamefont
  {Koizumi}, \citenamefont {Shin}, \citenamefont {Miyahara}, \citenamefont
  {Kimura}, \citenamefont {Miyamoto}, \citenamefont {Okuda}, \citenamefont
  {Namatame}, \citenamefont {Taniguchi}, \citenamefont {Arita}, \citenamefont
  {Nagaosa}, \citenamefont {Kobayashi}, \citenamefont {Murakami}, \citenamefont
  {Kumai}, \citenamefont {Kaneko}, \citenamefont {Onose},\ and\ \citenamefont
  {Tokura}}]{Ishizaka2011}%
  \BibitemOpen
  \bibfield  {author} {\bibinfo {author} {\bibfnamefont {K.}~\bibnamefont
  {Ishizaka}}, \bibinfo {author} {\bibfnamefont {M.~S.}\ \bibnamefont
  {Bahramy}}, \bibinfo {author} {\bibfnamefont {H.}~\bibnamefont {Murakawa}},
  \bibinfo {author} {\bibfnamefont {M.}~\bibnamefont {Sakano}}, \bibinfo
  {author} {\bibfnamefont {T.}~\bibnamefont {Shimojima}}, \bibinfo {author}
  {\bibfnamefont {T.}~\bibnamefont {Sonobe}}, \bibinfo {author} {\bibfnamefont
  {K.}~\bibnamefont {Koizumi}}, \bibinfo {author} {\bibfnamefont
  {S.}~\bibnamefont {Shin}}, \bibinfo {author} {\bibfnamefont {H.}~\bibnamefont
  {Miyahara}}, \bibinfo {author} {\bibfnamefont {A.}~\bibnamefont {Kimura}},
  \bibinfo {author} {\bibfnamefont {K.}~\bibnamefont {Miyamoto}}, \bibinfo
  {author} {\bibfnamefont {T.}~\bibnamefont {Okuda}}, \bibinfo {author}
  {\bibfnamefont {H.}~\bibnamefont {Namatame}}, \bibinfo {author}
  {\bibfnamefont {M.}~\bibnamefont {Taniguchi}}, \bibinfo {author}
  {\bibfnamefont {R.}~\bibnamefont {Arita}}, \bibinfo {author} {\bibfnamefont
  {N.}~\bibnamefont {Nagaosa}}, \bibinfo {author} {\bibfnamefont
  {K.}~\bibnamefont {Kobayashi}}, \bibinfo {author} {\bibfnamefont
  {Y.}~\bibnamefont {Murakami}}, \bibinfo {author} {\bibfnamefont
  {R.}~\bibnamefont {Kumai}}, \bibinfo {author} {\bibfnamefont
  {Y.}~\bibnamefont {Kaneko}}, \bibinfo {author} {\bibfnamefont
  {Y.}~\bibnamefont {Onose}}, \ and\ \bibinfo {author} {\bibfnamefont
  {Y.}~\bibnamefont {Tokura}},\ }\href {http://dx.doi.org/10.1038/nmat3051}
  {\bibfield  {journal} {\bibinfo  {journal} {Nat. Mater.}\ }\textbf {\bibinfo
  {volume} {10}},\ \bibinfo {pages} {521} (\bibinfo {year} {2011})}\BibitemShut
  {NoStop}%
\bibitem [{\citenamefont {Kato}\ \emph {et~al.}(2004)\citenamefont {Kato},
  \citenamefont {Myers}, \citenamefont {Gossard},\ and\ \citenamefont
  {Awschalom}}]{Kato2004}%
  \BibitemOpen
  \bibfield  {author} {\bibinfo {author} {\bibfnamefont {Y.~K.}\ \bibnamefont
  {Kato}}, \bibinfo {author} {\bibfnamefont {R.~C.}\ \bibnamefont {Myers}},
  \bibinfo {author} {\bibfnamefont {A.~C.}\ \bibnamefont {Gossard}}, \ and\
  \bibinfo {author} {\bibfnamefont {D.~D.}\ \bibnamefont {Awschalom}},\ }\href
  {\doibase 10.1126/science.1105514} {\bibfield  {journal} {\bibinfo  {journal}
  {Science}\ }\textbf {\bibinfo {volume} {306}},\ \bibinfo {pages} {1910}
  (\bibinfo {year} {2004})}\BibitemShut {NoStop}%
\bibitem{Kezsmarki2015}
I.~Kezsmarki, S.~Bordacs, P.~Milde, E.~Neuber, L.~M.~Eng, J.~S.~White, H.~M.~Ronnow,
C.~D.~Dewhurst, M.~Mochizuki, K.~Yanai, H.~Nakamura, D.~Ehlers, V.~Tsurkan, and A.~Loidl, Nature Materials, doi:10.1038/nmat4402 (2015).   
\end{thebibliography}

\end{document}